\documentclass{pasj01}
\usepackage{lmodern} % フォントサイズの警告を消すために追加
\usepackage[nopatch]{microtype}
\usepackage{booktabs}
\usepackage[switch,mathlines]{lineno}

\usepackage{etoolbox}

\makeatletter

\patchcmd{\@@maketitle}
  {\begingroup\historyfont
Received   \@make@formatted@date\rdate
 ; Accepted  \@make@formatted@date\adate\par\vspace*{8.5pt}%
\endgroup}
  {\vspace*{8.5pt}}
  {\typeout{SUCCESS: Received/Accepted removed}}
  {\typeout{ERROR: Received/Accepted patch failed}}

\def\ps@firstpage{%
   \def\@oddhead{%
      \hbox to \textwidth{\hfill{\foliofont\thepage}}%
   }%
   \let\@evenhead\@oddhead
   \def\@oddfoot{}%
   \def\@evenfoot{}%
   \let\@mkboth\@gobbletwo
}

\makeatother

\begin{document}
\makeatletter

\def\ps@headings{%
   \def\@evenhead{\vbox{\hsize\textwidth%
   {\evenfolio}\hfill\par\vspace*{5.5pt}
   {\vrule width\textwidth height1pt}}}%
   \def\@oddhead{\vbox{\hsize\textwidth%
   \hfill{\oddfolio}\par\vspace*{5.5pt}
   {\vrule width\textwidth height1pt}}}%
   \def\@oddfoot{\hfill}%
   \def\@evenfoot{\hfill}%
   \let\@mkboth\@gobbletwo
}

\pagestyle{headings}

\makeatother
\title{Probing the Three-Dimensional Structure of a Jet with Faraday Tomography}
\author{Sawera Gull\altaffilmark{1,*}}%
\author{Masaya Kurogi\altaffilmark{2}}%
\author{Keitaro Takahashi\altaffilmark{1}}%
\altaffiltext{1}{Kumamoto University, Graduate School of Science and Technology, 2-39-1, Kurokami, Chuo-ku, Kumamoto 860-8555, Japan}
\altaffiltext{2}{Department of R\&D, SoftService, Fukuoka, Japan}
\email{sawera.gull105@gmail.com}


\maketitle

\begin{abstract}
    Magnetic fields play a fundamental role in the dynamics and radiation of jets associated with active galactic nuclei (AGNs), where they accelerate and collimate the relativistic plasma and produce the observed synchrotron emission.
    Faraday tomography is a powerful technique for probing the three-dimensional distribution of magnetic fields and synchrotron-emitting plasma along the line of sight, by reconstructing the Faraday Dispersion Function (FDF) from the observed polarization spectrum.
    To explore what aspects of AGN-jet structure can be revealed by Faraday tomography, we construct a simple three-dimensional model of an AGN jet that captures its essential features, and compute the resulting FDF for a range of model parameters.
    The jet is represented as a cylindrical region of uniform thermal-electron density threaded by a coherent helical magnetic field, to which a random turbulent component can optionally be added.
    The parameters varied include the inclination angle $\theta$ between the jet axis and the line of sight, the wavenumber of the helical field, and the amplitude of the random magnetic-field component.
    By systematically varying these parameters, we examine how the underlying magnetic geometry and turbulence are encoded in the observable Faraday and polarization structures. The main observational signatures identified by our model include top-bottom asymmetry across the jet, curved or double peaked FDF profiles and increasingly fragmented Faraday depth structure along longer lines of sight, providing potential diagnostics of large scale helical magnetic fields and small scale turbulent components. These signatures providing a physical framework for interpreting future polarimetric observations of AGN jets from instruments such as LOFAR, MeerKAT, and the VLA.
\end{abstract}
%\pagewiselinenumbers
%\linenumbers

%%%%%%%%%%%%%%%%%%%%%%%%%%%%%%%%%%%%%%%%%%%%%%%%%%%%%%%%%%%%%%%%%%%%%%%
%%%%%%%%%%%%%%%%%%%%%%%%%%%%%%%%%%%%%%%%%%%%%%%%%%%%%%%%%%%%%%%%%%%%%%%
\section{Introduction}
%%%%%%%%%%%%%%%%%%%%%%%%%%%%%%%%%%%%%%%%%%%%%%%%%%%%%%%%%%%%%%%%%%%%%%%
%%%%%%%%%%%%%%%%%%%%%%%%%%%%%%%%%%%%%%%%%%%%%%%%%%%%%%%%%%%%%%%%%%%%%%%

Astrophysical jets associated with active galactic nuclei (AGNs) are high-velocity, collimated outflows of magnetized plasma that can extend from sub-parsec to Mpc scales.
They transport magnetic fields, kinetic energy, and heavy elements into the interstellar and intergalactic media, and are observed across the electromagnetic spectrum primarily through synchrotron emission.
Magnetic fields play a central role at every stage of jet evolution: they are believed to extract rotational energy from the central supermassive black hole or from its accretion disk through magnetohydrodynamic (MHD) mechanisms \citep{blandford1977electromagnetic,blandford1982hydromagnetic}, collimate the resulting outflow, and govern the synchrotron emission by which the jet is observed.
In such MHD-driven launching scenarios, the outflowing plasma is naturally threaded by a helical magnetic field whose poloidal and toroidal components evolve along the jet, so that the helical magnetic-field geometry serves as a key diagnostic of jet physics.

Observational evidence for large-scale helical magnetic fields in AGN jets has been steadily accumulating.
Transverse gradients in the rotation measure (RM) across the jet width, naturally produced by a toroidal field component, have been reported in many sources: \citet{goddi2025first} found large RMs and distinct RM gradients in submillimeter polarimetry of M87 and other AGNs, suggesting helical fields anchored near the jet base; \citet{toscano2025helical} reported a time-varying transverse RM gradient in 3C 273; \citet{livingston2025helical} found that strong RM and EVPA patterns associated with a helical toroidal field are concentrated in the innermost jet of NRAO 150; and \citet{peng2024faraday} traced a continuous helical RM gradient out to kpc scales in M87.
On the morphological side, \citet{orienti2024high} found a 90-degree rotation of the magnetic-field direction along the jet of PKS 1127$-$145, while \citet{anderson2022spiderweb} reported a sign reversal of the line-of-sight magnetic field across the jet width of the Spiderweb radio galaxy.
At smaller scales, deep VLBA polarimetry of blazar cores by \citet{kramer2025probing} and X-ray polarimetry of Mrk 421 by \citet{kim2024magnetic} both point to organized but evolving magnetic-field geometries close to the central engine.
Complementary observations of nearby radio galaxies, such as the head-tail source MRC 0600$-$399 studied with MeerKAT \citep{sakemi2025three}, and of the M87 jet itself \citep{park2026helical,pasetto2021reading}, reveal complex multi-component Faraday structures whose interpretation requires a fully three-dimensional view of the magnetic field.

A powerful technique for probing such three-dimensional magnetic-field structures along the line of sight is Faraday tomography.
The observed polarization spectrum $P(\lambda^2)$ is connected to the Faraday Dispersion Function (FDF) $F(\phi)$ through a Fourier transform pair \citep{burn1966depolarization}, where the Faraday depth $\phi$ encodes the path-integrated product of thermal electron density and line-of-sight magnetic field.
Each line of sight thus carries a one-dimensional tomographic profile of the magnetized plasma, and broadband multi-frequency polarimetry allows the reconstruction of $F(\phi)$ via the rotation measure synthesis technique introduced by \citet{brentjens2005faraday}.
A review of the theory and applications of Faraday tomography is given by \citet{2023PASJ...75S..50T}.
Recent methodological developments have substantially enhanced this capability: \citet{rudnick2024pseudo} demonstrated that polarized-intensity cube visualizations can separate foreground Faraday screens from emission local to the source, and \citet{gustafsson2025direction} reconstructed polarized emission with direction-dependent corrections to improve sensitivity to high Faraday depths.
The diagnostic power of Faraday tomography is expected to grow dramatically with the new generation of broadband polarimetric facilities such as LOFAR, MeerKAT, ASKAP, the VLA, and especially SKA1, whose wide and continuous frequency coverage will provide unprecedented Faraday-depth resolution and sensitivity, opening a new window onto the three-dimensional magnetic structure of AGN jets \citep{heald2020magnetism}.

Realistic AGN jets are unlikely to be purely coherent, however: small-scale random or turbulent magnetic-field components are expected to coexist with the large-scale helical field, and these turbulent components have direct observational signatures.
Using an RMHD-based polarization technique, \citet{jerrim2024faraday} demonstrated that synchrotron polarization in AGN jets is shaped jointly by turbulent magnetic fields and viewing-angle effects, and \citet{meenakshi2024comparative} showed that the strength, density, opening angle, and viewing geometry of AGN winds and jets all influence their radio and polarization signatures.
Existing analyses of polarization observations \citep{carrasco2010magnetized,pushkarev2023mojave} are typically interpreted using either simple analytic prescriptions or computationally expensive RMHD simulations \citep{baghel2024kpc,tsunetoe2025polarization,gelles2025signatures}, leaving a useful middle ground between these two extremes: a controlled, fully three-dimensional but inexpensive model that captures the essential geometric and magnetic ingredients of a helical jet and isolates how each parameter shapes the observed Faraday and polarization structure.

In this work, we develop such a simple three-dimensional model of an AGN jet and use it to examine what aspects of jet structure can be revealed by Faraday tomography.
The jet is represented as a cylindrical region of uniform thermal-electron density threaded by a coherent helical magnetic field of radius $a$, wavenumber $k_z$, and a prescribed ratio of perpendicular to parallel magnetic-field components $B_{\perp}/B_{\parallel}$, to which a random magnetic-field component can be added.
Using this model we compute the synchrotron emissivity, Faraday rotation, and polarization along each line of sight, and reconstruct the FDF by Fourier inversion.
We systematically vary the viewing angle $\theta$, the wavenumber $k_z$, and the amplitude of the random component in order to examine how these parameters are encoded in the observable Faraday and polarization structures.
We find that the FDF carries distinctive signatures of the helical magnetic-field geometry, that the viewing angle and wavenumber control the width and asymmetry of the FDF in characteristic ways, and that a random magnetic-field component preserves the underlying coherent morphology while fragmenting the FDF most strongly along the longest lines of sight.
This framework provides a physically transparent interpretation of Faraday and polarization observations of AGN jets, and complements approaches such as the polarized-cube studies of \citet{rudnick2024pseudo}, the three-dimensional Faraday tomographic analysis of \citet{sakemi2025three}, and the helical-field reconstructions of \citet{park2026helical}.

This paper is organized as follows.
In Section~\ref{sec:method}, we introduce the basics of Faraday rotation and Faraday tomography, and describe the three-dimensional helical jet model used in this study and the FDF calculation.
The results for the fiducial model and for cases with varying parameters are presented in Section~\ref{sec:results}, which is divided into four subsections.
In Section~\ref{sec:fiducial}, we present the FDF and polarization properties of the fiducial model.
In Section~\ref{sec:orientation}, we vary the orientation of the jet and examine its effect on the FDF.
In Section~\ref{sec:geometry}, we vary the wavenumber $k_z$ of the helical field to examine how the geometric structure of the jet is encoded in the FDF.
In Section~\ref{sec:random}, we add a random magnetic-field component and compare the results with the previous cases.
Finally, we discuss our findings and summarize the main conclusions in Section~\ref{sec:discussion}.

%%%%%%%%%%%%%%%%%%%%%%%%%%%%%%%%%%%%%%%%%%%%%%%%%%%%%%%%%%%%%%%%%%%%%%%
%%%%%%%%%%%%%%%%%%%%%%%%%%%%%%%%%%%%%%%%%%%%%%%%%%%%%%%%%%%%%%%%%%%%%%%
\section{Method}\label{sec:method}
%%%%%%%%%%%%%%%%%%%%%%%%%%%%%%%%%%%%%%%%%%%%%%%%%%%%%%%%%%%%%%%%%%%%%%%
%%%%%%%%%%%%%%%%%%%%%%%%%%%%%%%%%%%%%%%%%%%%%%%%%%%%%%%%%%%%%%%%%%%%%%%

%%%%%%%%%%%%%%%%%%%%%%%%%%%%%%%%%%%%%%%%%%%%%%%%%%%%%%%%%%%%%%%%%%%%%%%
\subsection{Faraday rotation}
%%%%%%%%%%%%%%%%%%%%%%%%%%%%%%%%%%%%%%%%%%%%%%%%%%%%%%%%%%%%%%%%%%%%%%%

The Stokes parameters $Q$ and $U$ can be used to express the polarization angle of the measured polarized emission.
\begin{equation}
    \chi = \frac{1}{2} \tan^{-1} \frac{U}{Q}
\end{equation}
The polarization plane of a polarized wave rotates in magnetized plasma, a phenomenon known as Faraday rotation.
It arises from the difference in the dispersion relation between left- and right-circularly polarized waves and the rotation angle of polarized radiation is proportional to the square of the wavelength and,
\begin{equation}
    \chi = \chi_0 + \textrm{RM} \lambda^2
\end{equation}
where $\lambda$ is the observing wavelength, $\chi_0$ is the intrinsic polarization angle, and RM is the rotation measure.
However, this is not sufficient for describing complicated magnetic-field structures, such as those found in individual galaxies and jets, because they include contributions from all regions with different values of RM along the line of sight.

%%%%%%%%%%%%%%%%%%%%%%%%%%%%%%%%%%%%%%%%%%%%%%%%%%%%%%%%%%%%%%%%%%%%%%%
\subsection{Basics of Faraday Tomography}
%%%%%%%%%%%%%%%%%%%%%%%%%%%%%%%%%%%%%%%%%%%%%%%%%%%%%%%%%%%%%%%%%%%%%%%

Faraday tomography is a technique for probing the structure of physical quantities, such as magnetic fields, thermal electron density, and polarized emission along the line of sight, from the observed polarization spectrum, $P(\lambda^2)$ (for a review, see \cite{2023PASJ...75S..50T}).
To describe the distribution of the polarized emission, we introduce the Faraday depth, $\phi$, which represents the degree to which the polarized emission at a specific location undergoes Faraday rotation.
Physically, $\phi$ is expressed as the integral of the magnetic-field component parallel to the line of sight and the thermal electron density along the line of sight:
\begin{eqnarray}
    && \phi = k \int_0^r n_e(r) B_{\parallel}(r) dr \\
    && k = \frac{e^3}{8\pi^2 \epsilon_0 m^2_e c^3}
\end{eqnarray}
where $k$ has a value of approximately $810~[{\rm rad/m^2}]$ when these parameters are expressed in the following units: $n_e~[\textrm{cm}^{-3}]$, $B_{\parallel}~[\mu \textrm{G}]$ and $dr~[\textrm{kpc}]$.
Using this variable, the Faraday dispersion function (FDF), $F(\phi)$, is defined as the complex polarized intensity distribution in $\phi$-space.
The observed polarization spectrum, $P(\lambda^2)$, can be expressed as the integral of the FDF over all Faraday depths:
\begin{equation}
    P(\lambda^2) = \int^{\infty}_{-\infty} F(\phi) e^{2i\phi \lambda^2} d\phi,
    \label{eq:PtoF}
\end{equation}
where the exponential factor represents the effect of Faraday rotation.
This equation mathematically takes the form of a Fourier transform, where the Faraday depth $\phi$ is the conjugate variable to the wavelength squared $\lambda^2$.
Formally, by applying the inverse Fourier transform, the FDF can be reconstructed from the observed polarization spectrum as follows:
\begin{equation}
    F(\phi) = \frac{1}{\pi} \int^{\infty}_{-\infty} P(\lambda^2) e^{-2i\phi \lambda^2} d\lambda^2.
    \label{eq:FtoP}
\end{equation}
In this way, the Faraday dispersion function $F(\phi)$, which encodes information about the line-of-sight distribution of physical quantities such as the magnetic field, thermal electron density, and polarized emission, can be derived from the observable $P(\lambda^2)$.
By performing this analysis along multiple lines of sight, one can effectively probe the three-dimensional structure of the astronomical object.

%%%%%%%%%%%%%%%%%%%%%%%%%%%%%%%%%%%%%%%%%%%%%%%%%%%%%%%%%%%%%%%%%%%%%%%
\subsection{Model of Helical Jet}
%%%%%%%%%%%%%%%%%%%%%%%%%%%%%%%%%%%%%%%%%%%%%%%%%%%%%%%%%%%%%%%%%%%%%%%

\begin{figure}[tb]
    \centering
    \includegraphics[width=0.72\linewidth]{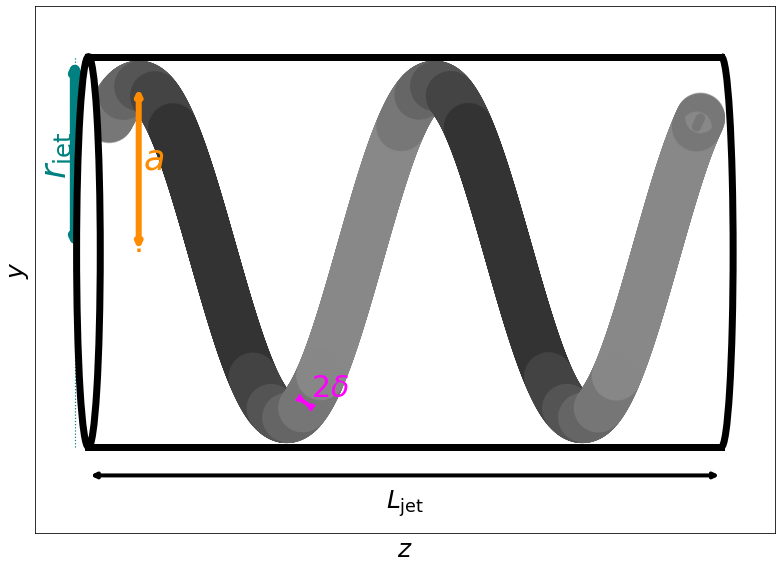}\\
    \vspace{0.35cm}
    \includegraphics[width=0.78\linewidth]{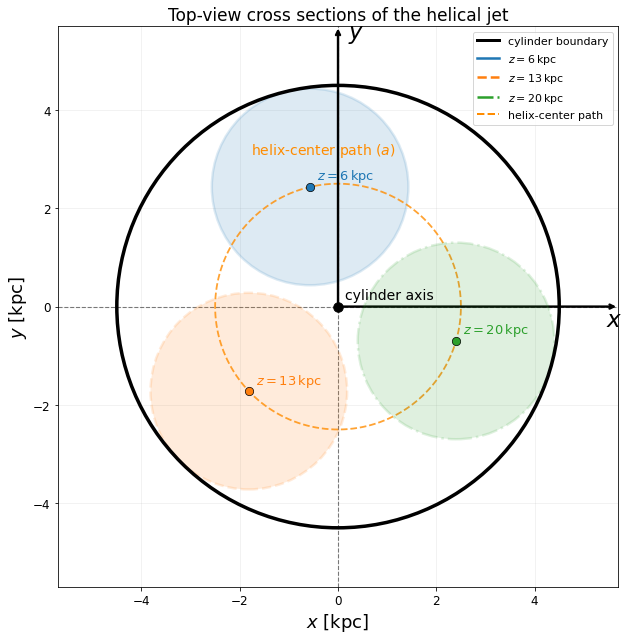}
    \caption{Geometry of the jet model.
    (a) Schematic illustration of the jet model containing a helical magnetic-field structure inside a cylinder.
    (b) Head-on cross sections of the cylinder, viewed down the jet ($z$) axis, at three representative positions $z=6$, $13$, and $20~\mathrm{kpc}$. The solid black circle is the cylinder boundary (radius $R_{\mathrm{cyl}}=4.5~\mathrm{kpc}$), the dashed orange circle is the helix-center path (radius $a=2.5~\mathrm{kpc}$), and the filled disks are the circular emitting cross sections (emitting-tube radius $r_{\mathrm{e}}=2.0~\mathrm{kpc}$) centered on the helix.\\
    Alt text: Cylindrical jet with helical magnetic field lines, and head-on cross sections of the cylinder.}
    \label{fig:jet-image}
\end{figure}

Figure \ref{fig:jet-image} illustrates our model: panel (a) is a schematic representation of the helical jet embedded in a cylinder, and panel (b) shows head-on cross sections of the cylinder (viewed down the jet axis) at three representative positions along the jet.
Here, $L_{\mathrm{jet}}$ denotes the length of the bounding cylinder (the jet length). The dashed circle of radius $a$ is the helix-center path, i.e., the radius at which the helix winds around the cylinder axis, and the emitting material occupies a tube of circular cross section of radius $r_{\mathrm{e}}$ centered on this helical path (the emitting-tube radius). The bounding cylinder therefore has radius $R_{\mathrm{cyl}} = a + r_{\mathrm{e}}$. We use these terms consistently throughout to avoid ambiguity between the outer cylinder and the emitting helical structure.
The cylinder is placed with its axis parallel to the z-axis and its center fixed at the Cartesian coordinate origin.
The integration along each line of sight is performed numerically on a Cartesian grid. The computational domain is discretized into $N_x \times N_y \times N_z = 200 \times 200 \times 200$ voxels, covering $x, y, z \in [-20,\,20]~\mathrm{kpc}$. For each line of sight, the Faraday depth is accumulated voxel by voxel. 

The center of the helix is parametrized as,
\begin{eqnarray}
    x &=& a \cos k_zz\\
    y &=& a \sin k_zz
\end{eqnarray}
where $k_z$ is the helix wavenumber.
We place the following magnetic field, which follows the helix, within a tubular region centered on a spiral curve satisfying the formula above.
\begin{eqnarray}
    B_x &=& - B_{\perp} \sin{k_z z}\\
    B_y &=& B_{\perp} \cos{k_z z}\\
    B_z &=& B_{\parallel}
\end{eqnarray}
The jet axis is aligned with the z-axis in the intrinsic coordinate system that defines the helical jet structure and its associated magnetic field.
Note that, because the magnetic field lines follow the helix, the relation $a k_z = B_{\perp}/B_{\parallel}$ holds.

To simulate different viewing geometries, both the helix and the magnetic-field vectors are rotated into the observer's frame.
The jet is first tilted with respect to the line of sight by applying a rotation $\theta$ about the y-axis.
The coordinate transformation that we use is
\begin{eqnarray}
    \mathbf{r}' &=& R_y(\theta) \mathbf{r}, \\
    \mathbf{B}' &=& R_y(\theta) \mathbf{B}
\end{eqnarray}
with
\begin{equation}
    R_y(\theta) =
    \left(
    \begin{array}{ccc}
        \cos\theta & 0 & \sin\theta\\
        0 & 1 & 0\\
        -\sin\theta & 0 & \cos\theta
    \end{array}
    \right).
\end{equation}
Thus, the transformed helix coordinates become, by substituting \(x=a\cos{k_z z}\) and \(y=a\sin{k_z z}\),
\begin{eqnarray}
    x' &=& a \cos{k_z z} \cos\theta + z\sin\theta, \\
    y' &=& a \sin{k_z z}, \\
    z' &=& -a\cos{k_z z} \sin\theta + z\cos\theta.
\end{eqnarray}
The transformed magnetic-field components are
\begin{eqnarray}
    B_x' &=& - B_{\perp} \sin{k_z z} \cos\theta + B_{\parallel} \sin\theta, \\
    B_y' &=& B_{\perp} \cos{k_z z}, \\
    B_z' &=& B_{\perp} \sin{k_z z} \sin\theta + B_{\parallel} \cos\theta.
\end{eqnarray}

Using Eqs.(18)-(20) the magnetic field components parallel and perpendicular to the line of sight are defined as 
\begin{equation}
    B_{\parallel, \mathrm{los}} = B'_z
\end{equation}

\begin{equation}
    B_{\perp,\mathrm{los}} = \sqrt{\left(B'_x\right)^2+
\left(B'_y\right)^2}
\end{equation}
Here, $B_{\parallel}$ and $B_{\perp}$ denote the intrinsic axial and transverse magnetic field components of the helical jet, whereas $B_{\parallel, \mathrm{los}}$ and $B_{\perp,\mathrm{los}}$ denote the magnetic field components parallel and perpendicular to the observer's line of sight.
The sign of the Faraday depth is determined by the sign of $B_{\parallel, \mathrm{los}}$. In our model, the viewing angle $\theta$ is defined as the angle between the jet axis and the observer's line of sight, such that $\theta = 0$ corresponds to the jet axis being aligned with the line of sight. For small viewing angles, the line of sight magnetic field remains predominantly positive. As the viewing angle increases, the projected magnetic field structure becomes more complex and $B_{\parallel, \mathrm{los}}$ can become negative in some regions of the jet, resulting in negative Faraday depths.
By combining the initial axial (z) and radial (x) components, this transformation changes the helix orientation and redistributes the magnetic-field components between the parallel and perpendicular directions with respect to the observer.
Both the magnetic field and the helix coordinates undergo the same linear transformation because they are treated as vectors.
Consequently, the line-of-sight component of the magnetic field, which drives Faraday rotation, and the observed geometry of the helix are modified consistently according to the selected viewing angle. We emphasize that the transformations in Equations~(14)--(22) are global, rigid rotations that do not depend on position within the jet, and are therefore applied identically to every voxel, both on and off the cylinder axis. This remains valid even though the emitting tube (radius $r_{\mathrm{e}}=2.0~\mathrm{kpc}$) occupies a large fraction of the cylinder (radius $R_{\mathrm{cyl}}=4.5~\mathrm{kpc}$) and extends well off the axis: the finite transverse extent of the emitting region is imposed separately, through the geometric condition that selects the emitting voxels, and does not alter the transformation itself.
In this case, the FDF can be computed from the polarization intensity, polarization angle, and Faraday depth, all of which can be determined from the magnetic-field components.
\begin{eqnarray}
    F(\phi) &=& \int \epsilon(x,z) e^{2i\chi(x,z)}\delta(\phi-\phi(x,z))d\textrm{z}\\
    \epsilon(x,z) &\propto& B_{\textrm{los}\perp}^{(\alpha+1)/2}
\end{eqnarray}
Here we take $\alpha = 3$, and the FDF intensity is proportional to $B_{\textrm{los}\perp}^{2}$. The parameters adopted in the fiducial jet model are summarized in Table~\ref{table}.
In our model, the magnetic field responsible for synchrotron radiation and Faraday rotation is present only within a finite-thickness region centered on the helix, so the entire cylinder does not radiate.
We consider a portion of the jet ejection region as a cylinder $40~\mathrm{kpc}$ in length and $9~\mathrm{kpc}$ in diameter, i.e., of radius $R_{\mathrm{cyl}}=4.5~\mathrm{kpc}$, which is the smallest cylinder that fully encloses the helical emitting region (helix-center radius $a=2.5~\mathrm{kpc}$ plus emitting-tube radius $r_{\mathrm{e}}=2.0~\mathrm{kpc}$).
Based on estimates from earlier studies by \cite{anderson2022spiderweb}, the thermal electron density and magnetic-field strength are fixed at 0.6 $[\textrm{cm}^{-3}]$ and 90 $[\mu \textrm{G}]$, respectively. The synthetic Faraday cubes presented in this work correspond to the intrinsic numerical resolution of the model, with no convolution by a finite telescope beam or rotation measure spread function. The computational domain spans $40~\mathrm{kpc}$ and is discretized into $200^3$ voxels, corresponding to a spatial resolution of $\sim0.2~\mathrm{kpc}$, and the Faraday dispersion function is sampled with 201 Faraday-depth channels.

It is useful to clarify which of our results depend on the absolute values of the model parameters and which do not. The morphology of the FDF in Faraday-depth space is governed by the line-of-sight profile of the product $n_e B_{\parallel}$ integrated along the path; rescaling $n_e$, the magnetic-field strength, or the physical size of the jet therefore only rescales the Faraday-depth ($\phi$) axis and leaves the shape of the FDF unchanged, so the geometric results are effectively scale-free. The polarization spectrum $P(\lambda^2)$ and the resulting depolarization, in contrast, depend on the actual Faraday depths sampled at a given wavelength through the factor $e^{2i\phi\lambda^2}$, and hence on the absolute values of $n_e$, $B_{\parallel}$, and the observing band. For this reason we adopt the observationally motivated values of \citet{anderson2022spiderweb} quoted above, these values are not intended to be representative of typical AGN jets but are specifically chosen to illustrate the Faraday signatures in an unusually high Faraday rotation environment. We evaluate $P(\lambda^2)$ over the range $\lambda^2 = 10^{-6}$--$10^{-3}~\mathrm{m}^2$ (Figures~\ref{fig:P-image} and \ref{fig:P-spectrum}).

\begin{table}[tb]
    \caption{Basic parameters and fiducial values}
    \centering
    \setlength{\tabcolsep}{3pt}
    \begin{tabular}{lcr} \hline
        Inclination angle & $\theta$ & $\pi/4$~[rad] \\ \hline
        Magnetic field strength & B & 90~[$\mu$G] \\ \hline
        Ratio of components & $B_{\perp}$/$B_{\parallel}$ & 0.5 \\ \hline
        Wave number & $k_z$ & $0.2~[\textrm{kpc}^{-1}]$ \\ \hline
        Thermal electron density & $n_e$ & $0.6~[\textrm{cm}^{-3}]$ \\ \hline
        Helix-center radius & $a$ & $2.5~[\textrm{kpc}]$ \\ \hline
        Emitting-tube radius & $r_{\rm e}$ & $2.0~[\textrm{kpc}]$ \\ \hline
        Cylinder radius & $R_{\rm cyl}$ & $4.5~[\textrm{kpc}]$ \\ \hline
    \end{tabular}
    \label{table}
\end{table}

%%%%%%%%%%%%%%%%%%%%%%%%%%%%%%%%%%%%%%%%%%%%%%%%%%%%%%%%%%%%%%%%%%%%%%%
%%%%%%%%%%%%%%%%%%%%%%%%%%%%%%%%%%%%%%%%%%%%%%%%%%%%%%%%%%%%%%%%%%%%%%%
\section{Results}\label{sec:results}
%%%%%%%%%%%%%%%%%%%%%%%%%%%%%%%%%%%%%%%%%%%%%%%%%%%%%%%%%%%%%%%%%%%%%%%
%%%%%%%%%%%%%%%%%%%%%%%%%%%%%%%%%%%%%%%%%%%%%%%%%%%%%%%%%%%%%%%%%%%%%%%

%%%%%%%%%%%%%%%%%%%%%%%%%%%%%%%%%%%%%%%%%%%%%%%%%%%%%%%%%%%%%%%%%%%%%%%
\subsection{Results for fiducial model}\label{sec:fiducial}
%%%%%%%%%%%%%%%%%%%%%%%%%%%%%%%%%%%%%%%%%%%%%%%%%%%%%%%%%%%%%%%%%%%%%%%

In this subsection, we discuss the properties of the fiducial model in detail, which is characterized by a helical magnetic-field structure embedded within a cylinder.
We first analyze the general behavior of the FDF and then focus on the detailed patterns in each region.

\begin{figure}[tb]
    \centering
    \includegraphics[width=0.4\textwidth]{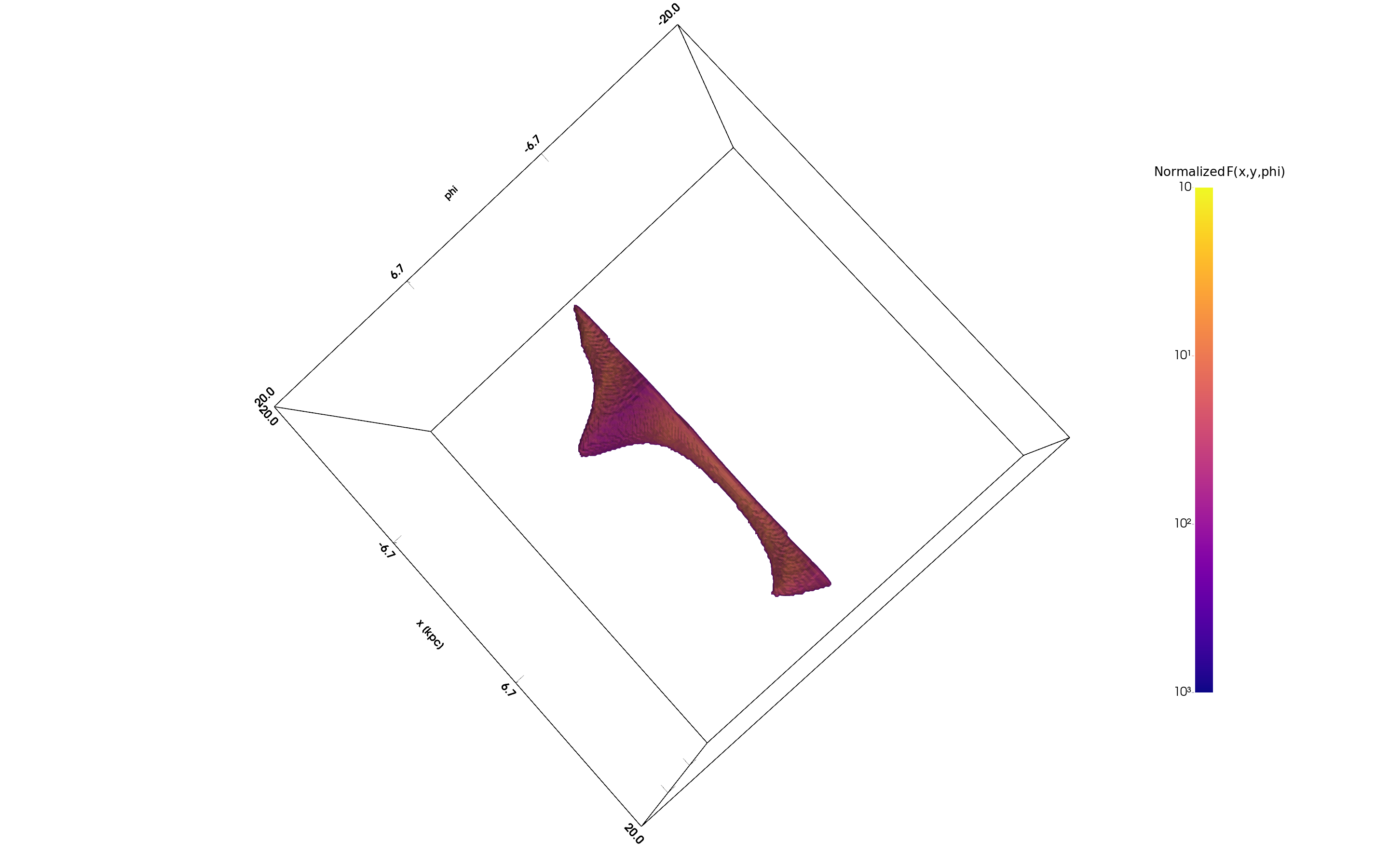}
    \vspace{0.4cm} % adjust spacing
    \includegraphics[width=0.4\textwidth]{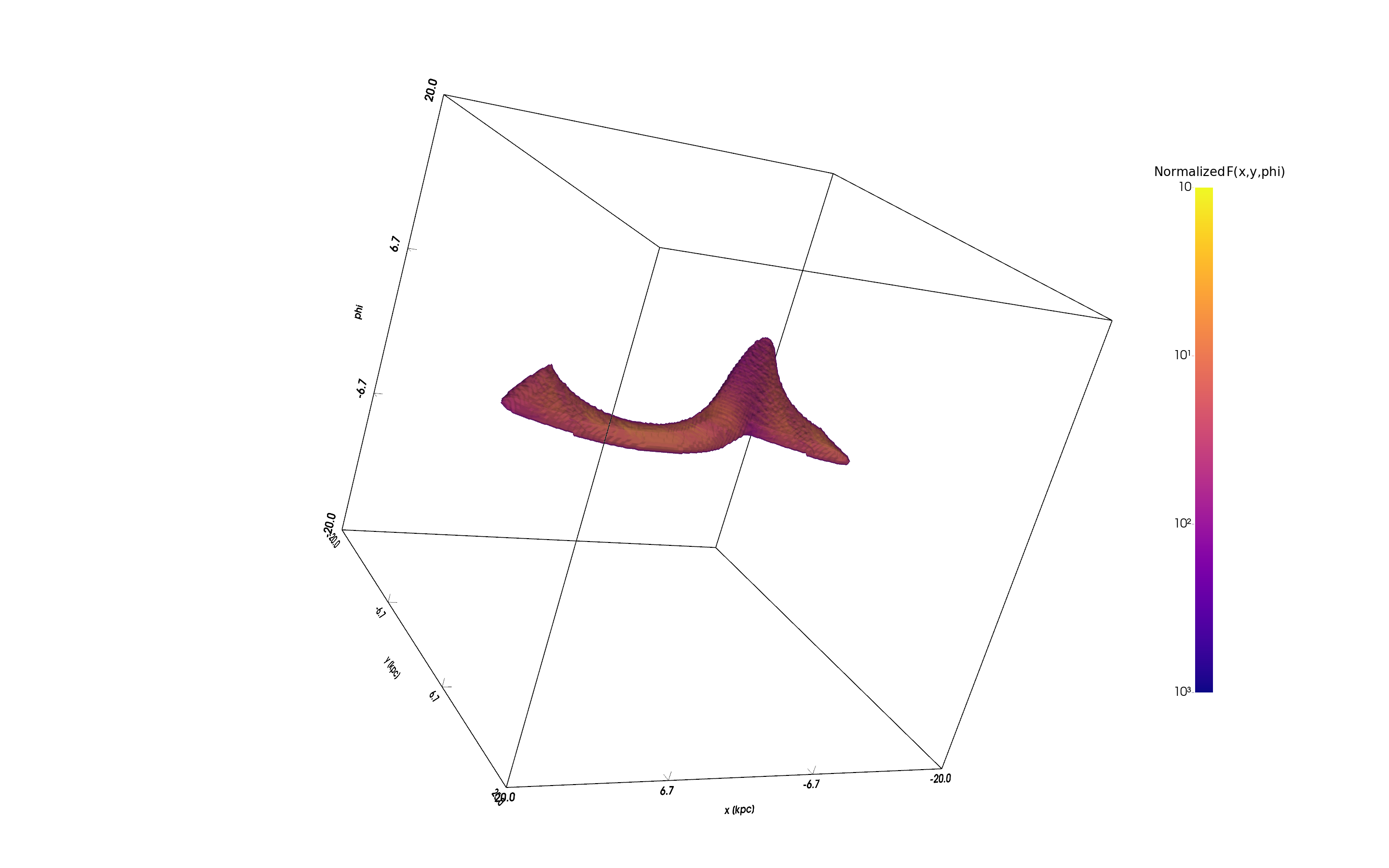}
    \caption{Three-dimensional iso-surface view of the cube $F(x,y,\phi)$ showing the distribution of polarized emission in Faraday depth $\phi$ across the jet for the fiducial model ($\theta=\pi/4$). The colors indicate the normalized FDF magnitude.\\
    Alt text: The three dimensional view of polarized emission in the jet.}
    \label{fig:cube}
\end{figure}

\begin{figure}[tb]
    \centering
    \includegraphics[width=0.7\linewidth]{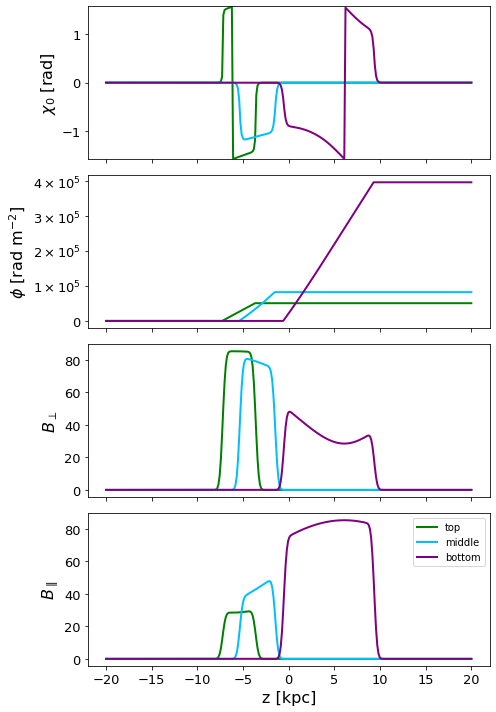}
    \caption{Profile of the magnetic-field components $B_{\parallel}$ and $B_{\perp}$, Faraday depth $\phi$, and intrinsic polarization angle $\chi_0$ from bottom to top, as functions of distance along the line of sight, for the fiducial model ($\theta=\pi/4$). The purple, light blue, and green lines correspond to the bottom, middle, and top of the helix, respectively.\\
    Alt text: Magnetic field , Faraday depth and polarization profiles along the jet.}
    \label{fig:profile}
\end{figure}

Figure \ref{fig:cube} illustrates the calculated three-dimensional distribution of polarized emission, where the spatial coordinates on the sky plane ($x, y$) are combined with the Faraday depth ($\phi$).
For visual clarity, the $\phi$ axis is rescaled to match the physical dimensions of the $x$ and $y$ axes, though the actual Faraday depth spans up to $\sim 10^{5}~[\textrm{rad m}^{-2}]$ (as detailed later in Figure \ref{fig:profile}).
In this idealized scenario without random turbulent magnetic fields, the polarized emission traces a distinctly continuous, smooth, ribbon-like structure in the $(x, y, \phi)$ space.
This unbroken curvature visually demonstrates how the line-of-sight component of the magnetic field ($B_{\parallel}$) varies systematically across the width and length of the jet.
Because the coherent helical magnetic field wraps tightly around the cylindrical jet, observing it at a viewing angle of $\theta = \pi/4$ creates regular spatial gradients and oscillations in $B_{\parallel}$.
These cyclic variations directly translate into the smooth undulating patterns along the $\phi$ axis, providing a symmetric Faraday structure characteristic of a large-scale ordered helical geometry.

\begin{figure}[tb]
    \centering
    \includegraphics[width=0.9\linewidth]{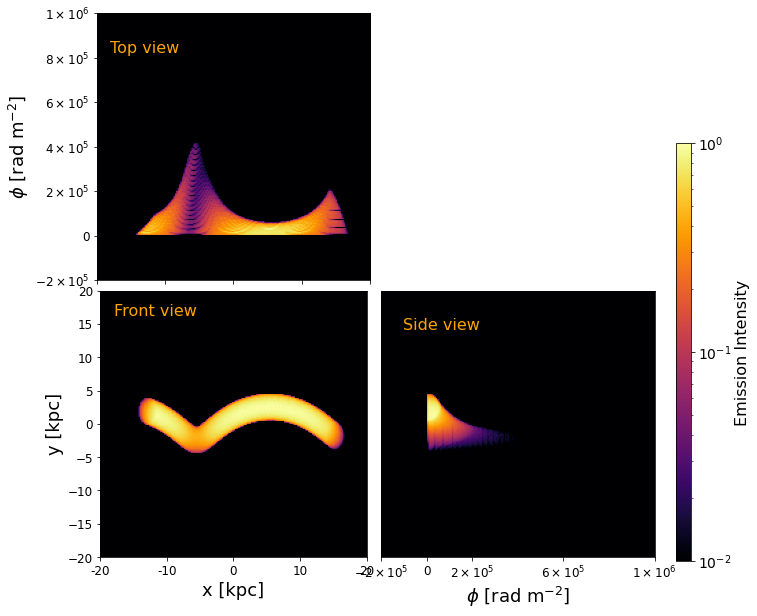}
    \caption{Three-view visualization of the projected synthetic emission and Faraday structure produced by the fiducial model ($\theta=\pi/4$).\\
    Alt text: Three projected views of the simulated jet emission and Faraday structure.}
    \label{fig:3views}
\end{figure}

Figure \ref{fig:3views} presents a three-view projection of the cube in Figure \ref{fig:cube}.
The bottom left panel shows the projection of the cube onto the x-y plane, defined as $F(x,y)=\int F(x,y, \phi)\,d\phi$; the bottom right panel shows the projection of the cube onto the $y$-$\phi$ plane, defined as $F(y,\phi)=\int F(x,y, \phi)dx$; and the top panel shows the projection of the cube onto the $x$-$\phi$ plane, defined as $F(x, \phi)=\int F(x,y, \phi)dy$.
The color scale represents the polarization intensity, which is proportional to the magnetic-field component perpendicular to the line of sight, because the electron density is uniform within the jet.
As stated before, in the fiducial model the jet morphology is determined solely by the coherent helical magnetic field, so the surface remains smooth and continuous and the jet spine, curvature, and symmetry are clearly visible.
The combination of magnetic-field components which are parallel and perpendicular to the line of sight causes the jet trajectory to oscillate smoothly.
The figure therefore provides a clear view of the axisymmetric geometry of the ordered helical jet.

\begin{figure}[tb]
    \centering
    \includegraphics[width=0.55\linewidth]{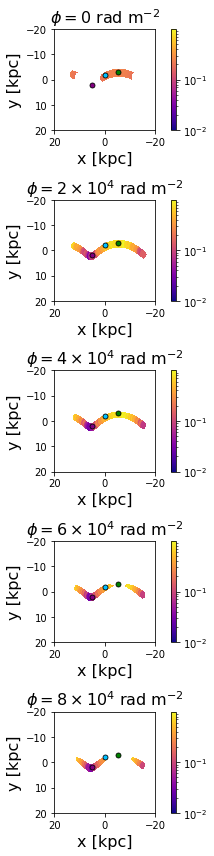}
    \caption{FDF map of a jet with a helical magnetic field for the fiducial model ($\theta=\pi/4$). The horizontal and vertical axes represent distance on the sky plane, and the color bar represents the FDF intensity in arbitrary units.\\
    Alt text: Maps of polarized emission across the simulated jet.}
    \label{fig:FDF-abs-image}
\end{figure}

Figure \ref{fig:FDF-abs-image} shows images of the jet at several values of $\phi$.
At $\phi = 0$, the emission is limited to compact regions near the central part and the top region near the green dot, rather than being distributed over the entire structure.
At $\phi = 2 \times 10^4$ and $4 \times 10^4~[\textrm{rad m}^{-2}]$, the emission extends over a larger fraction of the projected helix, including the bottom region near the purple dot, and forms a bright, curved ridge.
At higher Faraday depths, the ridge becomes less continuous: at $\phi = 6 \times 10^4~[\textrm{rad m}^{-2}]$, the emission around the top region near the green dot disappears, and at $\phi = 8 \times 10^4~[\textrm{rad m}^{-2}]$, the regions without detectable emission become larger.
This evolution shows that different parts of the helical jet dominate different Faraday-depth slices, producing a spatially fragmented FDF structure at large $\phi$.

\begin{figure}[tb]
    \centering
    \includegraphics[width=0.4\textwidth]{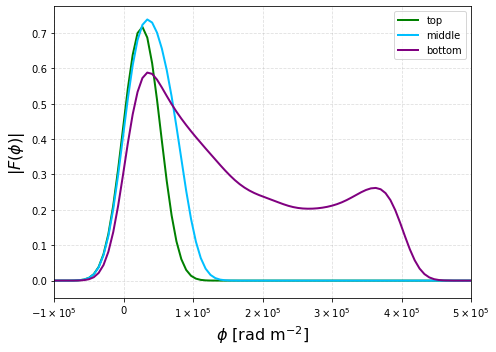}
    \includegraphics[width=0.4\textwidth]{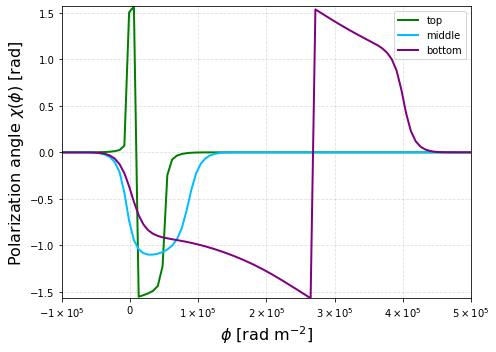}
    \vspace{0.4cm} % adjust spacing
    \caption{Absolute value of the FDF, $|F(\phi)|$ (top), and polarization angle $\chi(\phi)$ (bottom) along the line of sight corresponding to each position of the helix, for the fiducial model ($\theta=\pi/4$). The purple, light blue, and green lines correspond to the bottom, middle, and top of the helix, respectively.\\
    Alt text: FDF amplitude and polarization angle for three lines of sight.}
    \label{fig:FDF-abs-phase}
\end{figure}

Figure \ref{fig:FDF-abs-phase} presents the absolute value of the FDF in the upper panel and the polarization angle $\chi(\phi)$ in the lower panel for three representative points in the jet shown in Figure \ref{fig:FDF-abs-image}.
The green curve, which corresponds to the top of the jet, shows a sharp peak at around $\phi \sim 3 \times 10^4~[\textrm{rad m}^{-2}]$ and is confined to a relatively narrow range in Faraday depth.
The light-blue curve, corresponding to the middle of the jet, also remains concentrated in the positive-$\phi$ region.
In contrast, the purple curve from the bottom of the jet exhibits two smaller peaks, and the emission is distributed over a much broader range, extending to about $\phi \sim 4 \times 10^5~[\textrm{rad m}^{-2}]$.
This indicates that, for this parameter set, the magnetic-field component parallel to the line of sight is stronger in the bottom region, whereas the component perpendicular to the line of sight is stronger in the top region.
This top-bottom asymmetry arises from a geometric projection effect: because the helical magnetic field is tilted by $\theta = \pi/4$, the magnetic field lines in the top of the jet intersect the line of sight nearly perpendicularly, maximizing $B_{\perp}$.
Conversely, in the bottom, the field lines point more directly toward the observer, maximizing $B_{\parallel}$ and producing a broader FDF distribution.
The $\chi(\phi)$ panel shows that the green curve first becomes positive near the peak of the FDF and then changes to negative before returning to zero.
The light-blue curve is mainly negative over the Faraday-depth range where its absolute value is nonzero.
In contrast, the purple curve remains negative over a broad range of Faraday depth and then changes to positive at $\phi \sim 3 \times 10^5~[\textrm{rad m}^{-2}]$, indicating a qualitatively different Faraday structure in the bottom of the jet.

To understand the behavior of FDFs in Figure \ref{fig:FDF-abs-phase}, we show the profile of the magnetic field, Faraday depth, and intrinsic polarization angle as functions of physical distance along the line of sight, $z$, in Figure \ref{fig:profile}.
The bottom, middle, and top of the helical region are represented by the purple, light blue, and green lines, respectively.
Although the magnetic-field structure possesses helical symmetry, the line of sight is misaligned with the cylinder axis.
Consequently, the components perpendicular and parallel to the line of sight vary depending on the location within the jet, while the magnitude of the magnetic field is uniform everywhere.
In this fiducial model, the combination of the chosen pitch angle and the viewing angle ($\theta = \pi/4$) ensures that the line-of-sight magnetic field consistently points toward the observer (positive $B_{\parallel}$) so that the Faraday depth increases monotonically with distance along the line of sight in physical space, confining the FDF entirely to the positive $\phi$ region.
As we will discuss in Section~\ref{sec:orientation}, altering the viewing angle can introduce field reversals, causing the FDF to extend into negative $\phi$ values.
Therefore, in the fiducial model, there is a one-to-one correspondence between $z$ and $\phi$, which makes the interpretation of the FDF straightforward.
The polarized intensity is correlated with the magnetic-field component perpendicular to the line of sight, while the intrinsic polarization angle $\chi_0$ is determined by the direction of magnetic field projected onto the sky.
By examining the Faraday-depth profile together with the profile of the perpendicular magnetic-field component, we can understand the behavior of the FDF, that is, polarized intensity as a function of $\phi$.
In this fiducial model, the magnetic-field component parallel to the line of sight has a step-like profile.
Therefore, the Faraday depth not only increases monotonically, but also increases almost in proportion to the distance within the jet.
As a result, the shape of the FDF closely resembles the profile of the magnetic-field component perpendicular to the line of sight.

\begin{figure}[tb]
    \centering
    \includegraphics[width=0.95\linewidth]{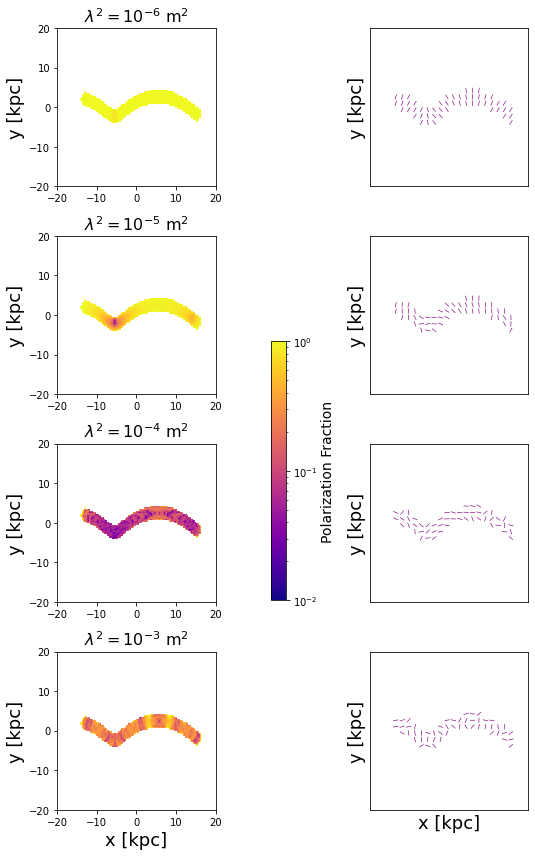}
    \caption{Maps of the polarization fraction (left) and polarization angle (right) for each value of wavelength squared, for the fiducial model ($\theta=\pi/4$). The color bar indicates the polarization fraction.\\
    Alt text: Polarization fraction and angle maps at different wavelength.}
    \label{fig:P-image}
\end{figure}

By Fourier transforming the FDF, we obtain the polarization spectrum, $P(\lambda^2)$, computed here over the range $\lambda^2 = 10^{-6}$--$10^{-3}~\mathrm{m}^2$, which is the directly observable quantity.
Figure \ref{fig:P-image} shows the polarization fraction (left) and the vector map of the polarization angle (right) at several wavelengths.
In general, Faraday rotation becomes more significant at longer wavelengths.
Because the degree of Faraday rotation varies across different locations within the jet, the superposition of emission originating from these regions along a single line of sight results in depolarization.
Consequently, the polarization fraction decreases as the wavelength increases.
Furthermore, because the depolarization process depends on the magnetic-field structure along the line of sight, the spatial distribution of the polarization fraction is non-uniform.
Specifically, depolarization is more pronounced in the bottom of the jet, which is directly related to the broader extent of the FDF in $\phi$-space in this region, as we saw in Figure \ref{fig:FDF-abs-phase}.

The right panels present the vector maps of the observed polarization angles.
At short wavelengths, where Faraday rotation is negligible, the polarization angle clearly reflects the intrinsic magnetic-field structure of the jet.
Conversely, at longer wavelengths, the observed emission is a superposition of components that have experienced varying degrees of Faraday rotation along the line of sight.
As a result, the observed polarization angle is determined by the weighted sum of these differently rotated polarization planes.
As seen in the figure, the polarization angles become increasingly disordered across the jet at longer wavelengths.
Such top-bottom asymmetry in the polarization fraction and FDF profile across the jet cross-section serves as a crucial observational signature for identifying large-scale helical magnetic fields in AGN jets.

\begin{figure}[tb]
    \centering
    \includegraphics[width=0.7\linewidth]{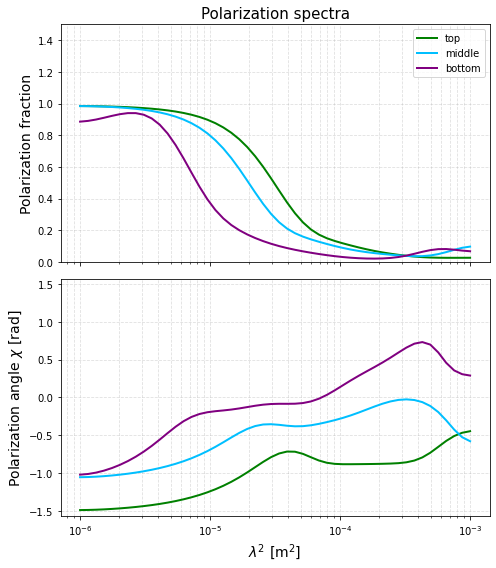}
    \caption{Polarization spectra $P(\lambda^2)$ for the fiducial model ($\theta=\pi/4$): absolute values (top) and polarization angle (bottom). The purple, light blue and green lines correspond to the bottom, middle and top of the helix, respectively, indicated in Fig.~\ref{fig:FDF-abs-image}.\\
    Alt text: Polarization intensity and angle versus wavelength squared.}
    \label{fig:P-spectrum}
\end{figure}

Figure \ref{fig:P-spectrum} illustrates the absolute values (top) and phases (bottom) of the polarization spectra for the bottom (purple line), middle (light blue line), and top (green line) of the helix.
As shown in the top panel, the polarized intensity decays at longer wavelengths for all locations.
Notably, this decay occurs at much shorter wavelengths in the bottom of the jet.
This rapid depolarization is a direct consequence of the FDF in the bottom having a larger spread in $\phi$-space, as previously observed in Figure \ref{fig:FDF-abs-phase}.
The bottom panel demonstrates that at short wavelengths, the polarization angle remains almost constant because the effect of Faraday rotation is negligible.
However, as the wavelength increases, the polarization angle exhibits continuous variations due to the increasing effect of Faraday rotation.

%%%%%%%%%%%%%%%%%%%%%%%%%%%%%%%%%%%%%%%%%%%%%%%%%%%%%%%%%%%%%%%%%%%%%%%
\subsection{Orientation of the Jet}\label{sec:orientation}
%%%%%%%%%%%%%%%%%%%%%%%%%%%%%%%%%%%%%%%%%%%%%%%%%%%%%%%%%%%%%%%%%%%%%%%

We next vary the inclination angle to $\pi/8$ and $3\pi/8$ to examine how the FDF shape changes relative to the fiducial model with $\theta=\pi/4$.
As in the fiducial model, the top and bottom of the helix behave differently because the magnetic-field components parallel and perpendicular to the line of sight vary with position in the jet.
Changing the inclination angle modifies both the path length through the magnetized region and the projection of the helical magnetic field onto the line of sight. The two limiting geometries, the pole-on case ($\theta=0$) and the plane-of-sky (edge-on) case ($\theta=\pi/2$), are presented in Appendix~\ref{sec:appendix}; the plane-of-sky case yields FDF and polarization-angle distributions that are symmetric in the sign of $\phi$ and $\chi$, and provides a useful reference for the intermediate inclinations considered here.

\begin{figure}[tb]
    \centering
    \includegraphics[width=0.9\linewidth]{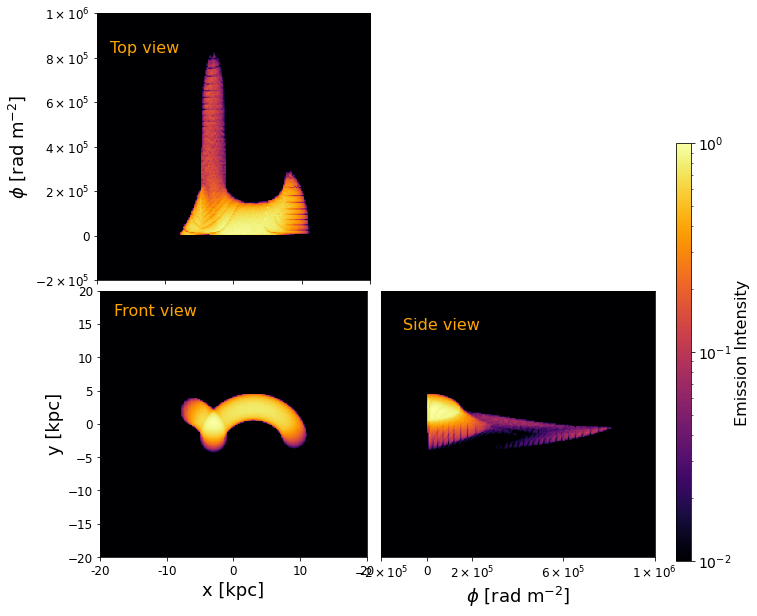}
    \includegraphics[width=0.9\linewidth]{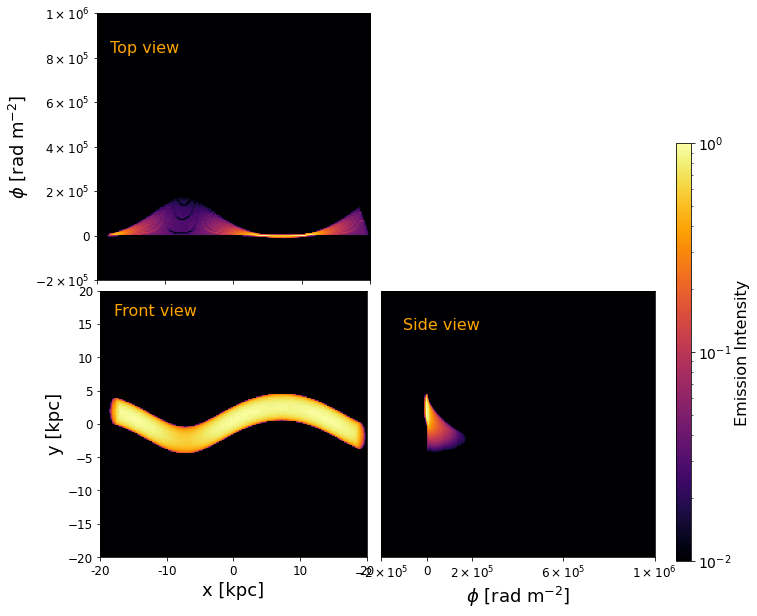}
    \caption{Three-view visualization same as Figure \ref{fig:3views}, but for inclination angles $\theta=\pi/8$ (top) and $\theta=3\pi/8$ (bottom).\\
    Alt text: Projected jet emission for two inclination angles.}
    \label{fig:3views-inclination}
\end{figure}

Figure \ref{fig:3views-inclination} presents the same three-view projection as Figure \ref{fig:3views}, but for different inclination angles.
For $\theta=\pi/8$, the projected emitting region is more compact on the sky plane than in the fiducial model, while the distribution in Faraday depth extends to larger positive $\phi$.
This high-Faraday-depth emission appears where different parts of the helical emitting region overlap along the line of sight.
For $\theta=3\pi/8$, the projected helix is more extended on the sky plane, but the emission is concentrated closer to $\phi=0$ and the overall Faraday-depth distribution becomes narrower.
In this case, a small part of the emission extends to negative $\phi$, indicating that the line-of-sight component of the magnetic field changes sign in part of the jet. The $\theta = 3\pi/8$ viewing angle is close to, but not identical to the plane of sky, leaving a finite line of sight magnetic field component. Consequently, the accumulated Faraday depth is slightly asymmetric, resulting in predominantly positive Faraday depth values with weaker negative components. The exactly symmetric limit ($\theta=\pi/2$), in which the positive and negative Faraday-depth contributions cancel, is shown for comparison in Appendix~\ref{sec:appendix}.

\begin{figure}[t]
    \begin{tabular}{cc}
        \begin{minipage}[t]{0.5\hsize}
            \centering
            \includegraphics[width=0.9\linewidth]{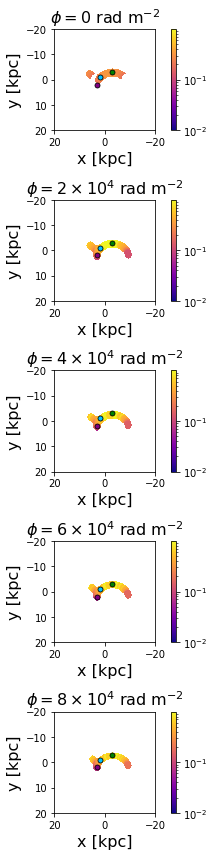}
        \end{minipage}
        \begin{minipage}[t]{0.5\hsize}
            \centering
            \includegraphics[width=0.9\linewidth]{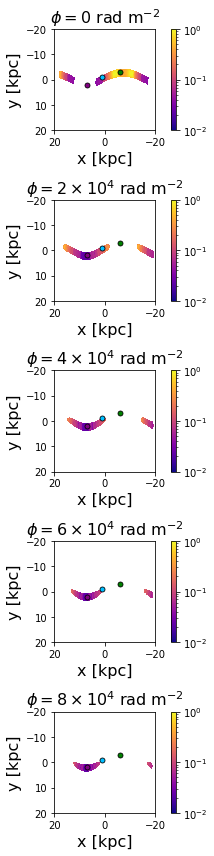}
        \end{minipage}
    \end{tabular}
    \caption{FDF map of a jet with a helical magnetic field same as Figure \ref{fig:FDF-abs-image}, but for inclination angles $\theta=\pi/8$ (left) and $\theta=3\pi/8$ (right).\\
    Alt text: FDF maps for two different jet inclination angles.}
    \label{fig:FDF-abs-image-inclination}
\end{figure}

\begin{figure}[ht]
    \centering
    \includegraphics[width=0.9\linewidth]{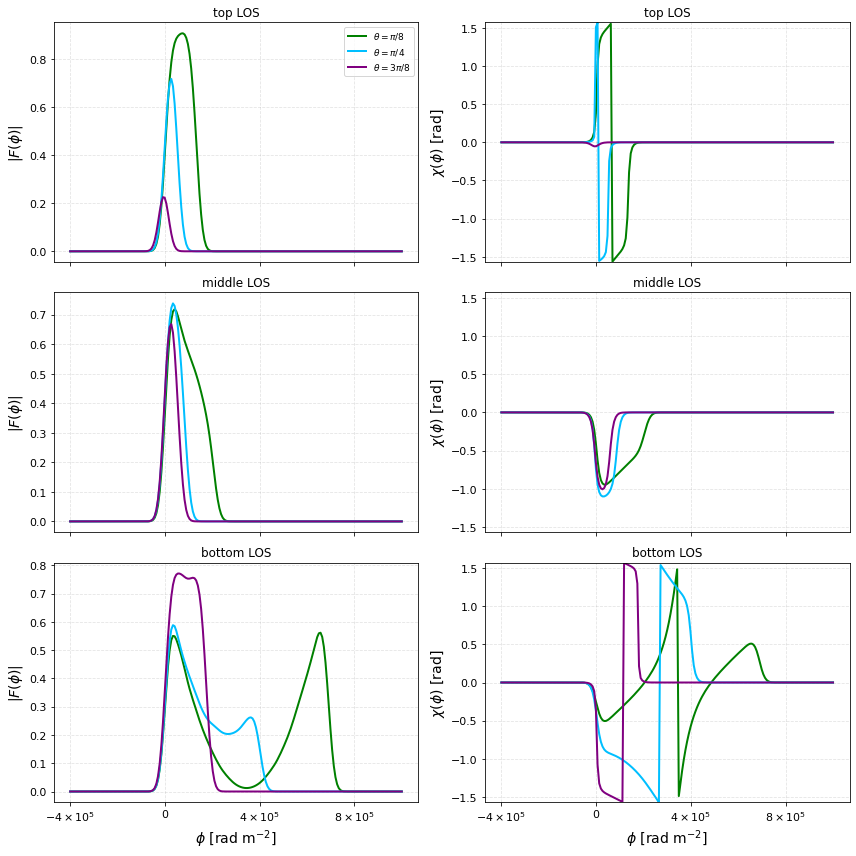}
    \caption{Comparison of the absolute value of the FDF (left column) and the polarization angle (right column) for different lines of sight through the helix. The top, middle, and bottom rows correspond to the top, middle, and bottom of the helix, respectively. The inclination angles are $\theta=\pi/8$ (green line), $\theta=\pi/4$ (light blue line), and $\theta=3\pi/8$ (purple line).\\
    Alt text: Multiple line plots comparing the FDF amplitude and polarization angle for different jet inclinations.}
    \label{fig:FDF-comparison}
\end{figure}

Figure \ref{fig:FDF-abs-image-inclination} shows the FDF maps at the same Faraday-depth slices as Figure \ref{fig:FDF-abs-image}.
For $\theta=\pi/8$, emission remains visible over much of the projected ridge even at $\phi=6\times10^4$ and $8\times10^4~[\textrm{rad m}^{-2}]$, unlike the fiducial model where the high-$\phi$ maps become more fragmented.
This indicates that a smaller inclination angle produces a broader FDF distribution because the line of sight passes through a longer magnetized path and intersects overlapping parts of the helix.
For $\theta=3\pi/8$, the emission is prominent near $\phi=0$, but it becomes increasingly localized as $\phi$ increases.
Thus, compared with the fiducial model, the larger inclination angle concentrates the polarized emission into a narrower range of Faraday depth.

Figure \ref{fig:FDF-comparison} shows the FDF (left column) and polarization angle $\chi(\phi)$ (right column) along three lines of sight.
The inclination angles $\pi/8$, $\pi/4$, and $3\pi/8$ are represented by the green, light blue, and purple lines, respectively, and the top, middle, and bottom rows correspond to the top, middle, and bottom of the helix.
At the top and middle of the helix, the FDF peak shifts to smaller Faraday depth as the inclination angle increases from $\pi/8$ to $3\pi/8$.
This trend is consistent with Figures \ref{fig:3views-inclination} and \ref{fig:FDF-abs-image-inclination}, where the Faraday-depth distribution becomes narrower at larger inclination angle.
At the bottom of the helix, the difference from the fiducial model is more pronounced: the $\theta=\pi/8$ curve shows a clearly separated high-$\phi$ component, whereas the fiducial curve also has two peaks but the emission between them remains relatively strong, so the two peaks are not well separated.
The $\theta=3\pi/8$ curve is concentrated at lower Faraday depth.
The polarization-angle profiles show the same projection effect through their amplitudes and oscillatory patterns.
At the top and middle of the helix, the amplitude of the polarization-angle variation becomes smaller as the inclination angle increases.
At the bottom of the helix, the $\theta=\pi/8$ and fiducial curves show multiple changes in polarization angle, whereas the $\theta=3\pi/8$ curve is dominated by a single sharp swing from negative to positive values.
Because Faraday depth and polarized intensity are related to the magnetic-field components parallel and perpendicular to the line of sight, respectively, these differences show that changing the inclination angle alters the direction of the magnetic field relative to the line of sight.

\begin{figure}[th]
    \centering
    \begin{tabular}{cc}
        \begin{minipage}[t]{0.5\hsize}
            \centering
            \includegraphics[width=0.9\linewidth]{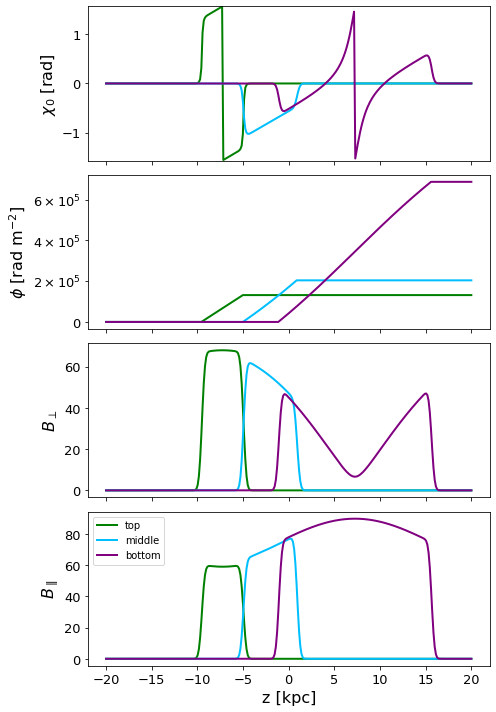}
        \end{minipage}
        \begin{minipage}[t]{0.5\hsize}
            \centering
            \includegraphics[width=0.9\linewidth]{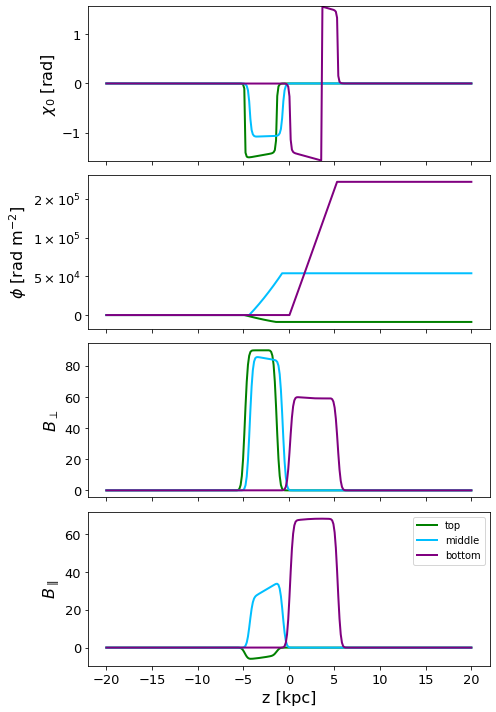}
        \end{minipage}
    \end{tabular}
    \caption{Profile of the magnetic-field components $B_{\parallel}$ and $B_{\perp}$, Faraday depth $\phi$, and intrinsic polarization angle $\chi_0$ same as Figure \ref{fig:profile}, but for inclination angles $\theta=\pi/8$ (left) and $\theta=3\pi/8$ (right).\\
    Alt text: Line plots showing magnetic field strength, Faraday depth profiles for two jet inclinations.}
    \label{fig:profile-inclination}
\end{figure}

Figure \ref{fig:profile-inclination} confirms this interpretation in physical space along the line of sight.
For $\theta=\pi/8$, the line-of-sight magnetic-field component remains positive for the three representative positions, so the Faraday depth increases monotonically and reaches large positive values, especially at the bottom of the helix.
The peaks of the magnetic-field component perpendicular to the line of sight are also spread over a longer distance in real space, which produces the separated high-$\phi$ component seen in Figure \ref{fig:FDF-comparison}.
For $\theta=3\pi/8$, the magnetized region traversed by the line of sight becomes narrower, and the upper line of sight includes a negative $B_{\parallel}$ component.
As a result, the FDF is compressed into a smaller range of $\phi$, while a small negative Faraday-depth component appears in the top of the helix.

%%%%%%%%%%%%%%%%%%%%%%%%%%%%%%%%%%%%%%%%%%%%%%%%%%%%%%%%%%%%%%%%%%%%%%%
\subsection{Geometric Structure of the Jet}\label{sec:geometry}
%%%%%%%%%%%%%%%%%%%%%%%%%%%%%%%%%%%%%%%%%%%%%%%%%%%%%%%%%%%%%%%%%%%%%%%

\begin{figure}[tb]
    \centering
    \includegraphics[width=0.9\linewidth]{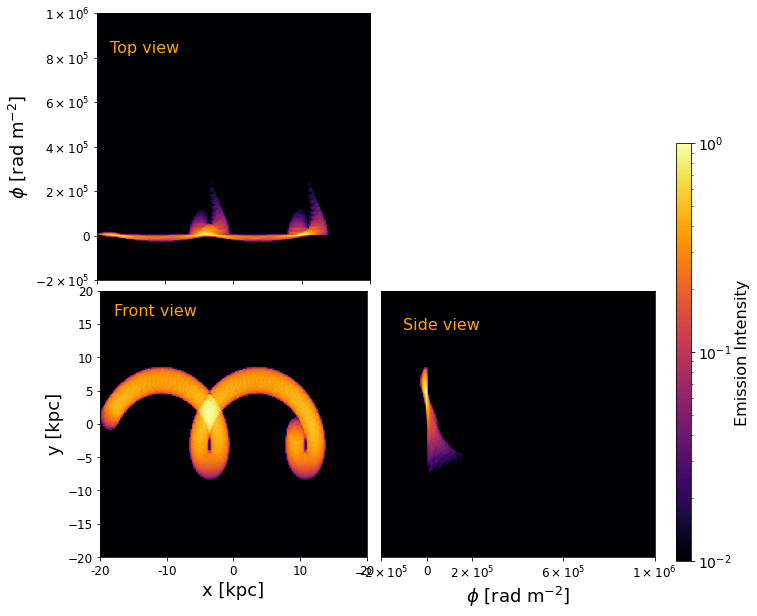}
    \caption{Three-view visualization same as Figure \ref{fig:3views}, but for $k_z = 0.8~{\rm kpc}^{-1}$, at the fiducial viewing angle $\theta=\pi/4$.\\
    Alt text: Projected jet emission for a larger parameter of wave number.}
    \label{fig:3views-B2}
\end{figure}

In comparison to the previous subsections, we now examine our model with an increased wavenumber ($k_z = 0.8~{\rm kpc}^{-1}$ and, accordingly, $B_{\perp}/B_{\parallel}= 2.0$) to observe how it varies from our fiducial model.
Figure \ref{fig:3views-B2} shows the same three-view visualization of the synthetic emission and Faraday structure as in Figure \ref{fig:3views}, but for the increased wavenumber.
For $k_z = 0.8~{\rm kpc}^{-1}$, the distribution of Faraday depth shrinks to smaller positive $\phi$, with a small region of emission extending to negative $\phi$, and the emission is concentrated close to $\phi=0$.
This negative-$\phi$ extension arises because, at higher $k_z$, the line-of-sight magnetic field $B_{z'} = B_{\perp}\sin{k_z z} \sin\theta + B_{\parallel}\cos\theta$ oscillates with a larger relative amplitude (since $B_{\perp}$ now dominates over $B_{\parallel}$), causing local sign reversals along the path that produce a small negative-$\phi$ contribution.

\begin{figure}[tb]
    \centering
    \includegraphics[width=0.55\linewidth]{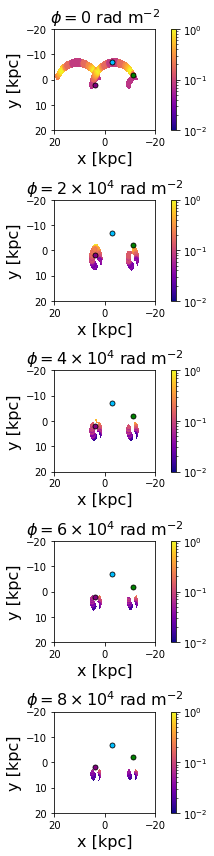}
    \caption{FDF map of a jet with a helical magnetic field same as Figure \ref{fig:FDF-abs-image}, but for $k_z = 0.8~{\rm kpc}^{-1}$. The viewing angle is $\theta=\pi/4$.\\
    Alt text: FDF maps for a larger parameter of wave number.}
    \label{fig:FDF-abs-image-B2}
\end{figure}

The FDF map in Figure \ref{fig:FDF-abs-image-B2} illustrates the magnetic field arrangement for the case of $k_z = 0.8~{\rm kpc}^{-1}$.
Figure \ref{fig:FDF-abs-image-B2} is the same as Figure \ref{fig:FDF-abs-image}, showing the same Faraday-depth slices, but the emission becomes increasingly localized as $\phi$ increases, and more prominent emission appears near $\phi=0$.
At $\phi=0$, the emission is extended and traces approximately two full helical wavelengths across the projected jet, in contrast to the single wavelength visible in the fiducial model (Figure \ref{fig:FDF-abs-image}), reflecting the fourfold increase in $k_z$.
As $\phi$ increases, the emission becomes separated into compact regions, showing fragmented behavior similar to that in Figure \ref{fig:FDF-abs-image}.
Faraday-depth slices at $\phi=2\times10^4$ and $4\times10^4~[\textrm{rad m}^{-2}]$ are significantly splintered as compared to the fiducial model, which showed clear apparent emission over a large portion of the projected slice.

\begin{figure}[tb]
    \centering
    \includegraphics[width=0.9\linewidth]{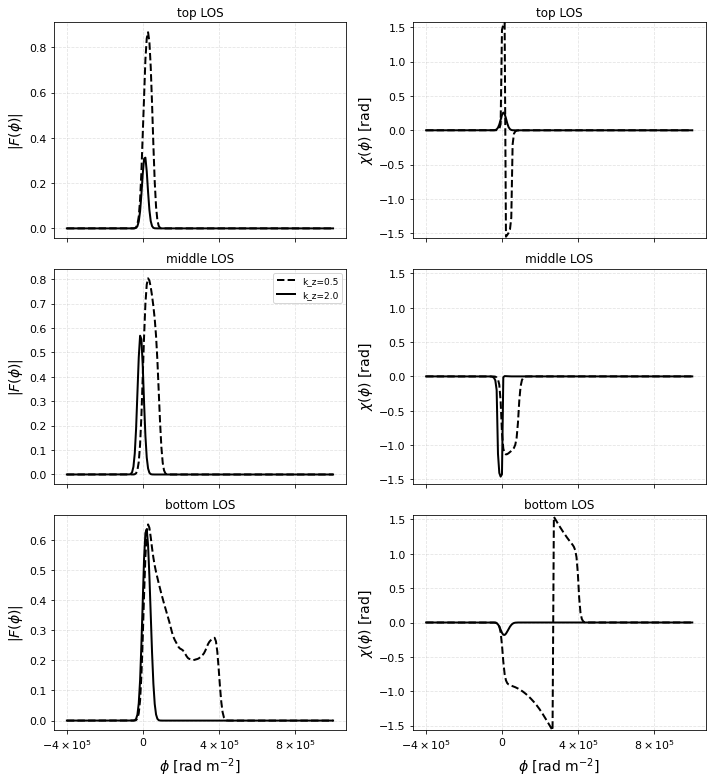}
    \caption{Comparison of the absolute value of the FDF (left column) and the polarization angle (right column) same as Figure \ref{fig:FDF-comparison} but for comparison between $k_z = 0.2~{\rm kpc}^{-1}$ and $k_z = 0.8~{\rm kpc}^{-1}$, both at the fiducial viewing angle $\theta=\pi/4$.\\
    Alt text: Multiple line plots comparing FDF properties for two different helical wave number parameters.}
    \label{fig:FDF-abs-phase-B2} 
\end{figure}

Figure \ref{fig:FDF-abs-phase-B2} illustrates the FDF (left column) and polarization angle $\chi(\phi)$ (right column) along three lines of sight for two distinct wavenumbers.
The solid line corresponds to $k_z = 0.8~{\rm kpc}^{-1}$, while the dashed line corresponds to $k_z = 0.2~{\rm kpc}^{-1}$.
For both cases at the top and bottom of the helix, the Faraday depth starts near $\phi=0$ and extends to positive $\phi$, although it is evident that $k_z = 0.2~{\rm kpc}^{-1}$ extends to higher $\phi$-values, while $k_z = 0.8~{\rm kpc}^{-1}$ terminates earlier.
At the middle line of sight of the helix, the solid line starts at a lower Faraday depth than the dashed line, showing a small negative Faraday-depth component.
The polarization angle $\chi(\phi)\,[\textrm{rad}]$ changes sharply near the locations where $|F(\phi)|$ has strong peaks.
These rapid angle variations show that the polarized signal is sensitive to the internal helical structure.

Thus, in comparison to the fiducial model, the FDF for $k_z = 0.8~{\rm kpc}^{-1}$ generally exhibits smaller peaks (most clearly at the top and middle lines of sight, while at the bottom line of sight the peak heights are comparable) and the emission extends over a shorter range, about $\phi \sim 1 \times 10^5~[\textrm{rad m}^{-2}]$, whereas for $k_z = 0.2~{\rm kpc}^{-1}$ the jet exhibits higher peaks and the emission is distributed over a much broader range, about $\phi \sim 4 \times 10^5~[\textrm{rad m}^{-2}]$.
From this case, we conclude that a smaller wavenumber enhances the Faraday-depth complexity and broadens the polarized emission in $\phi$-space, whereas a larger wavenumber produces more compact and coherent Faraday spectra.
A greater number of helical turns can be contained within the given computational domain of 40 kpc by increasing the helical wavenumber, which reduces the helical wavelength. As a result, a single line of sight may cross several rounds of the helical magnetic field before exiting the cylinder, depending on the viewing angle. A wider and more intricate Faraday dispersion function is produced by each intersection's contribution at a distinct Faraday depth. Conversely, lower values of $k_z$ result in simpler Faraday-depth structures because they have fewer helical turns within the same physical length. As a result, the observed Faraday spectra are dependent on the magnetic-field configuration as well as the viewing geometry and the number of helical cycles contained in the computational domain.

%%%%%%%%%%%%%%%%%%%%%%%%%%%%%%%%%%%%%%%%%%%%%%%%%%%%%%%%%%%%%%%%%%%%%%%
\subsection{Random Magnetic Field}\label{sec:random}
%%%%%%%%%%%%%%%%%%%%%%%%%%%%%%%%%%%%%%%%%%%%%%%%%%%%%%%%%%%%%%%%%%%%%%%

In this subsection, we investigate the effect of random magnetic fields on the FDF structure.
To approximate turbulent fields, we add normally distributed fluctuations with standard deviation $B_{\rm random}=90~[\mu {\rm G}]$, equal to the magnitude of the coherent field, to each magnetic-field component ($B_x$, $B_y$, $B_z$) at every grid point, superposed on the coherent helical field, while other parameters are fixed to the fiducial values.

\begin{figure}[tb]
    \centering
    \includegraphics[width=0.8\linewidth]{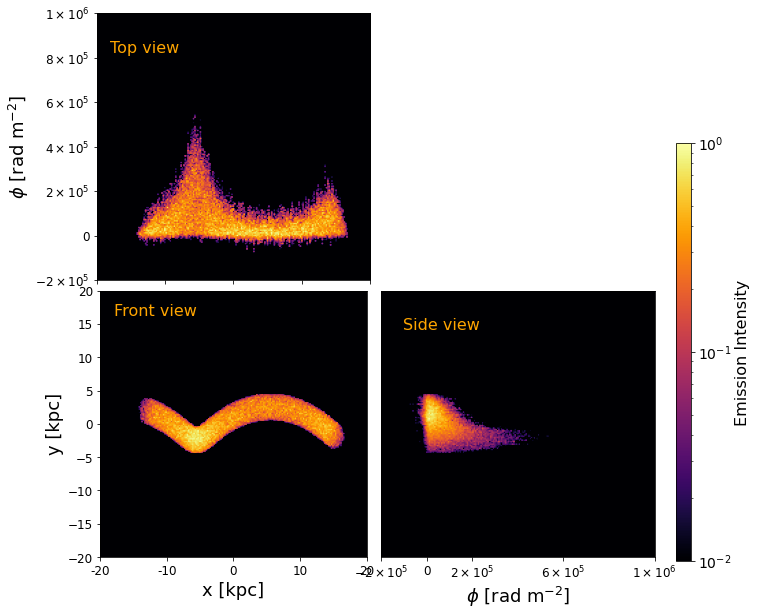}
    \caption{Three-view visualization same as Figure \ref{fig:3views}, but with a random magnetic field of $B_{\rm random}=90~[\mu {\rm G}]$. The viewing angle is the fiducial $\theta=\pi/4$.\\
    Alt text: Projected jet emission with an added random magnetic field component.}
    \label{fig:3views-Brand}
\end{figure}

Figure \ref{fig:3views-Brand} illustrates the same three-view projection as Figure \ref{fig:3views}, but with the random magnetic field.
In the front view, the large-scale helical morphology is preserved, since it is set by the coherent component, while the emission within the helical spine becomes mottled, reflecting local fluctuations in $B_{\perp}^2$.
In the top view, the smooth ribbon-like structure of the fiducial model is replaced by a scattered distribution spreading over a broader range of Faraday depth, with isolated bright pixels reaching $\phi \sim 5 \times 10^5~[\textrm{rad m}^{-2}]$ near the upper end of the helix (compared with the smooth peak at $\sim 3 \times 10^5$ in the fiducial model), and a small fraction of emission extending into negative $\phi$.
The side view similarly shows that the originally compact distribution near $\phi=0$ becomes broadened and noisy.
These changes arise because random fluctuations in $B_{z'}$ contribute stochastically to the Faraday-rotation integrand, enhancing the dispersion of $\phi$ along each line of sight and occasionally producing local sign reversals that drive $\phi$ to negative values.
A turbulent field of amplitude comparable to the coherent one is thus sufficient to disrupt the coherent Faraday structure without erasing its underlying geometric form.

\begin{figure}[tb]
    \centering
    \includegraphics[width=0.5\linewidth]{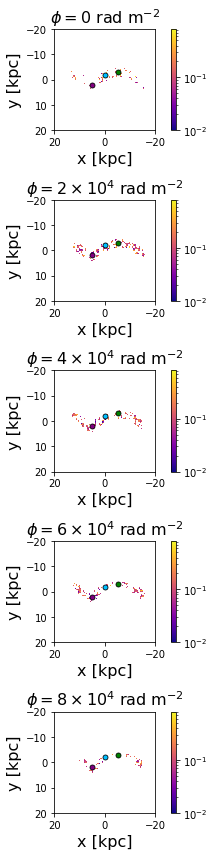}
    \caption{FDF map of a jet with a helical magnetic field same as Figure \ref{fig:FDF-abs-image}, but with a random magnetic field of $B_{\rm random}=90~[\mu {\rm G}]$. The viewing angle is $\theta=\pi/4$.\\
    Alt text: Two dimensional FDF maps with added random magnetic field component.}
    \label{fig:FDF-abs-image-Brand}
\end{figure}

Figure \ref{fig:FDF-abs-image-Brand} shows images of the jet at several values of $\phi$, as in Figure \ref{fig:FDF-abs-image}, but with the random magnetic field included.
At $\phi = 0$, the compact coherent emission of the fiducial model is now scattered into many small, disconnected pixels distributed along the projected helical path.
At $\phi = 2 \times 10^4$ and $4 \times 10^4~[\textrm{rad m}^{-2}]$, the emission is similarly fragmented and no longer traces a continuous curved ridge, although the bottom region near the purple dot retains relatively more emission.
At $\phi = 6 \times 10^4$ and $8 \times 10^4~[\textrm{rad m}^{-2}]$, the fragmentation worsens and extended regions near the green dot show little detectable emission.
The overall envelope of bright pixels still follows the projected helical path, but the coherent ridge-like features of the fiducial model are no longer present, because random fluctuations in $B_{z'}$ cause the local Faraday depth at each pixel to deviate stochastically from its coherent value, redistributing emission across neighboring $\phi$-slices.

\begin{figure}[tb]
    \centering
    \includegraphics[width=0.8\linewidth]{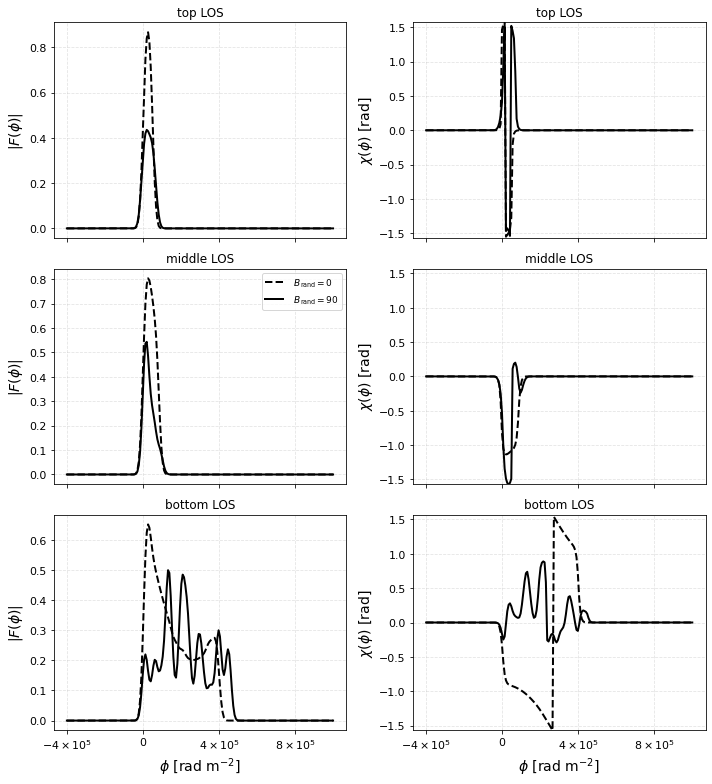}
    \caption{Absolute value of the FDF, $|F(\phi)|$ (left), and polarization angle $\chi(\phi)$ (right) along the line of sight same as Figure \ref{fig:FDF-abs-phase} for the fiducial model, comparing the cases without and with a random magnetic field of $B_{\rm random}=90~[\mu {\rm G}]$. The viewing angle is $\theta=\pi/4$.\\
    Alt text: Line plots comparing the FDF properties with and without random magnetic field component.}
    \label{fig:FDF-abs-phase-Brand}
\end{figure}

Figure \ref{fig:FDF-abs-phase-Brand} compares the FDF (left column) and polarization angle $\chi(\phi)$ (right column) along the three lines of sight for the fiducial model without (dashed line) and with (solid line) the random magnetic field.
In the top and middle lines of sight, the inclusion of the random field reduces the peak amplitude of $|F(\phi)|$ (from $\sim 0.85$ to $\sim 0.4$ and from $\sim 0.8$ to $\sim 0.55$, respectively) and slightly broadens the distribution, but the overall single-peak structure near $\phi \approx 0$ is preserved and $\chi(\phi)$ is only modestly perturbed.
The bottom line of sight, by contrast, shows a qualitative change: the originally smooth distribution extending to $\phi \sim 5 \times 10^5~[\textrm{rad m}^{-2}]$ is fragmented into a series of comparably bright sub-peaks of amplitude $\sim 0.2$--$0.5$, and $\chi(\phi)$ develops pronounced oscillations across the full $\phi$-range.
This contrast reflects the coupling between turbulence and the coherent FDF: where the coherent distribution is intrinsically narrow (top and middle), the random fluctuations have a limited $\phi$-range over which to redistribute emission, whereas where it is broad (bottom), the same fluctuations disperse the emission across many independent $\phi$-slices.

\begin{figure}[tb]
    \centering
    \includegraphics[width=0.8\linewidth]{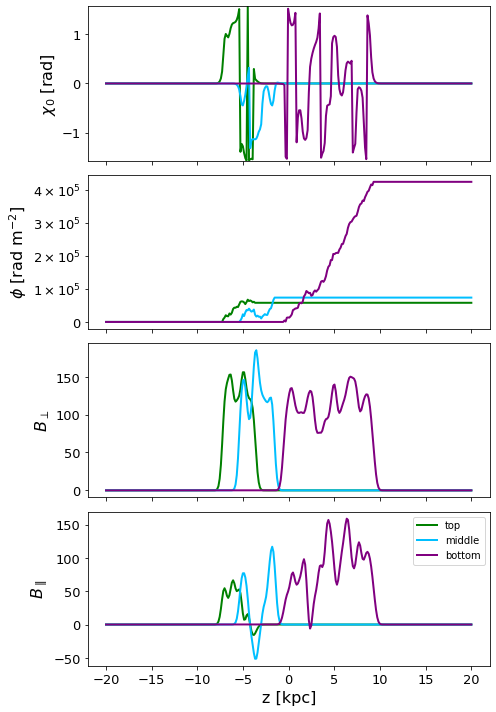}
    \caption{Profile of the magnetic-field components $B_{\parallel}$ and $B_{\perp}$, Faraday depth $\phi$, and intrinsic polarization angle $\chi_0$ same as Figure \ref{fig:profile}, but for $B_{\rm random}=90~[\mu {\rm G}]$. The viewing angle is $\theta=\pi/4$.\\
    Alt text: Line plots showing magnetic field strength, Faraday profiles with a random magnetic field component.}
    \label{fig:profile-Brand}
\end{figure}

Figure \ref{fig:profile-Brand} shows the line-of-sight profiles of $\chi_0$, $\phi$, $B_{\perp}$, and $B_{\parallel}$ as functions of $z$ for the three lines of sight, as in Figure \ref{fig:profile}, but with the random magnetic field.
$B_{\perp}$ and $B_{\parallel}$ now display multiple sub-peaks superposed on the coherent envelope.
For the top line of sight, which intersects only a short segment of the helix, the overall shape of the profiles is essentially intact, with $\phi$ rising smoothly to its plateau and $\chi_0$ only modestly perturbed.
The middle line of sight, although also short, shows the strongest local effect: $B_{\parallel}$ undergoes a sharp negative excursion reaching $\sim -50~[\mu {\rm G}]$ around $z \sim -2$ kpc, which causes $\phi$ to decrease locally over this range before plateauing at $\sim 0.7 \times 10^5~[\textrm{rad m}^{-2}]$, accompanied by rapid oscillations of $\chi_0$ in the same region.
The bottom line of sight traverses the jet over a much longer path ($\Delta z \sim 10$ kpc); along this path $B_{\parallel}$ breaks into multiple positive sub-peaks with only minor negative excursions, $B_{\perp}$ similarly develops multiple sub-peaks, and $\chi_0$ oscillates rapidly between $\pm \pi/2$ across the entire interval.
Even so, $\phi$ still rises monotonically to $\sim 4 \times 10^5~[\textrm{rad m}^{-2}]$, indicating that the coherent gradient dominates the integrated Faraday depth despite the strong local fluctuations.
Because $B_{\parallel}$ enters the Faraday-rotation integrand and $\chi_0$ accumulates over the full path, the bottom line of sight is the most strongly affected by turbulence in the cumulative sense, which explains its fragmented $|F(\phi)|$ and oscillatory $\chi(\phi)$ in Figure \ref{fig:FDF-abs-phase-Brand}.
This case resembles the morphology of real AGN jets where coherent and turbulent magnetic-field components coexist, consistent with observations in which complex or multiple FDF peaks indicate regions of turbulent magnetic structure or mixed emitting zones.

%%%%%%%%%%%%%%%%%%%%%%%%%%%%%%%%%%%%%%%%%%%%%%%%%%%%%%%%%%%%%%%%%%%%%%%
%%%%%%%%%%%%%%%%%%%%%%%%%%%%%%%%%%%%%%%%%%%%%%%%%%%%%%%%%%%%%%%%%%%%%%%
\section{Discussion and Summary}\label{sec:discussion}
%%%%%%%%%%%%%%%%%%%%%%%%%%%%%%%%%%%%%%%%%%%%%%%%%%%%%%%%%%%%%%%%%%%%%%%
%%%%%%%%%%%%%%%%%%%%%%%%%%%%%%%%%%%%%%%%%%%%%%%%%%%%%%%%%%%%%%%%%%%%%%%

In this paper, we constructed a simple model of an AGN jet, which consists of a cylindrical region containing thermal electrons at a uniform density, threaded by a coherent helical magnetic field, and computed its Faraday Dispersion Function (FDF) under various parameter choices.
We varied the inclination angle of the jet, the wavenumber of the helical magnetic field, and the amplitude of a superposed random magnetic-field component, in order to examine how the observed polarization structure encodes the internal three-dimensional geometry of the magnetic field.
Compared with two-dimensional models, this fully three-dimensional treatment incorporates depth-dependent variations in both emission and magnetic-field direction along each line of sight, which is essential for properly characterizing the Faraday complexities of magnetized AGN jets.

For the fiducial model ($\theta = \pi/4$, $B_{\perp}/B_{\parallel} = 0.5$, $k_z = 0.2~{\rm kpc}^{-1}$), the polarized emission traces a smooth, ribbon-like surface in $(x, y, \phi)$ space that reflects the coherent helical geometry.
Because $B_{\parallel}$ remains positive across the jet, $\phi$ increases monotonically along the line of sight, confining the FDF entirely to positive $\phi$, and $|F(\phi)|$ closely resembles the $B_{\perp}$ profile.
The geometric projection produces a clear top-bottom asymmetry: $B_{\parallel}$ dominates at the bottom, yielding a broader FDF extending to $\phi \sim 4 \times 10^5~[\textrm{rad m}^{-2}]$, while $B_{\perp}$ dominates at the top, yielding a sharper peak; this also leads to stronger depolarization at the bottom.

Varying the parameters produces systematic changes in the FDF.
A smaller inclination angle ($\theta = \pi/8$) broadens the FDF and produces a clearly separated high-$\phi$ component at the bottom, whereas a larger one ($\theta = 3\pi/8$) compresses the FDF and introduces a small negative-$\phi$ component from local sign reversals of $B_{\parallel}$ at the top.
Increasing the wavenumber to $k_z = 0.8~{\rm kpc}^{-1}$ ($B_{\perp}/B_{\parallel} = 2.0$) yields a more compact FDF confined to $\phi \lesssim 1 \times 10^5~[\textrm{rad m}^{-2}]$, again with a small negative-$\phi$ extension from the larger oscillation of $B_{z'}$.
Adding a random magnetic-field component with amplitude comparable to the coherent field preserves the helical morphology but fragments the FDF: the top and middle lines of sight are only modestly affected, while the bottom one, which has the longest path through the jet, develops multiple sub-peaks in $|F(\phi)|$ and rapid oscillations in $\chi(\phi)$ from accumulated stochastic fluctuations.

These model results have direct implications for the interpretation of polarimetric observations of AGN jets.
The diagnostic morphologies identified above (top-bottom asymmetry, curved or double-peaked FDF profiles, and fragmented spectra along long lines of sight) provide concrete observational signatures of large-scale helical magnetic fields and the coexistence of small-scale turbulent components.
By directly comparing simulated and observed FDFs along multiple lines of sight, one can constrain parameters such as jet orientation, helix pitch, and the relative strength of coherent and random magnetic fields, complementing approaches such as the polarized-cube studies of \citet{rudnick2024pseudo} that aim to separate intrinsic jet magnetization from foreground Faraday screens.
Multi-frequency polarimetric data from high-resolution surveys such as LOFAR, MeerKAT, and the VLA are particularly well suited for this comparison, and may allow the reconstruction of FDFs spanning long jet structures.

The present model relies on several simplifying assumptions that should be kept in mind when interpreting the results.
The jet is treated as a cylinder of fixed radius and length, with both thermal electrons and synchrotron-emitting relativistic electrons distributed at uniform density inside the cylinder.
The magnetic field has uniform magnitude that only varies in direction, with the coherent component fixed to a strict helical geometry.
The random magnetic-field component is modeled as Gaussian fluctuations added independently to each Cartesian component, which is a crude approximation to genuine magnetohydrodynamic turbulence.
Relativistic effects on the jet emission and beaming are neglected, and the jet is treated as a static structure without time evolution.
These idealizations are useful for isolating the dependence of the FDF on the key geometric and magnetic parameters, but a quantitative comparison with observed AGN jets will eventually require refining the electron density and magnetic-field distributions inside the cylinder and incorporating information about the surrounding plasma.

In the next phase of this research, the simulated FDFs from the 3D model will be directly compared with those obtained from actual radio observations of AGN jets, with the aim of verifying whether the modeled helical field and its Faraday-depth variations reproduce the observed spectra and spatial signatures and of constraining parameters such as jet orientation, pitch angle, and field-strength distribution.
This framework will be extended by incorporating synthetic observation effects such as beam convolution and noise, allowing direct comparison with high-resolution polarimetric data from instruments such as LOFAR, ASKAP, and MeerKAT.
For comparison, a typical VLA observation at $1.4~\mathrm{GHz}$ with a synthesized beam of $\sim1.3^{\prime\prime}$ corresponds to a physical resolution of approximately $0.6~\mathrm{kpc}$ for a source at a distance of $100~\mathrm{Mpc}$. Consequently, the fine-scale Faraday structures predicted by the present model would be partially smoothed by the telescope beam, while the finite RMSF would broaden nearby Faraday-depth components. Therefore, the present results should be regarded as the intrinsic Faraday structure prior to instrumental convolution.
The present model assumes that synchrotron emission and Faraday rotation arise within the same cylindrical plasma, and therefore does not include an external, magnetized Faraday-rotating sheath surrounding the jet. This is a deliberate simplification whose limitations deserve emphasis. In sources such as the Spiderweb radio galaxy \citep{anderson2022spiderweb}, on which our fiducial parameters are based, a surrounding sheath can provide a Faraday-rotation contribution comparable in magnitude to that of the jet body itself (except toward the very brightest knots), so it is a first-order effect rather than a small correction. Such a sheath would add a largely position-dependent Faraday-depth offset and gradient along each line of sight, which could partially mask, enhance, or even mimic the intrinsic signatures identified above. We nonetheless adopt the single-medium description here because our goal is to isolate the Faraday and polarization signatures produced by the intrinsic helical field of the emitting plasma, without the additional degeneracy introduced by a separate external screen. Incorporating an external Faraday-rotating sheath, and disentangling its contribution from that of the internal field, is therefore an important extension that we leave to future work.
We also plan to develop a statistical framework for measuring the degree of agreement between observed and synthesized FDFs, applying methods such as RM synthesis, QU fitting, and Faraday tomography inversion consistently to both simulated and actual data cubes.
The ultimate goal of this work is to improve our knowledge of the magnetic-field structure in AGN jets by bridging the gap between theoretical jet magnetohydrodynamic simulations and polarimetric radio measurements.

\section*{Acknowledgment}
KT is partially supported by JSPS KAKENHI grant Nos. 24H01813, 25K21670, 26H00838 and 26K21724.

\appendix
\section{Limiting viewing geometries: pole-on and plane-of-sky}\label{sec:appendix}

For completeness, and to orient the reader with respect to the intermediate inclinations discussed in Section~\ref{sec:orientation}, we present here the two limiting viewing geometries of the fiducial model: the pole-on case, in which the jet axis is aligned with the line of sight ($\theta=0$), and the plane-of-sky (edge-on) case, in which the jet axis lies in the plane of the sky ($\theta=\pi/2$). All other parameters are identical to the fiducial model (Table~\ref{table}). The plane-of-sky case is particularly relevant for large-scale jets on tens-to-hundreds-of-kpc scales, which are typically viewed at large angles to the line of sight, in contrast to the small viewing angles characteristic of VLBI-scale jets.

\begin{figure}[tb]
    \centering
    \begin{tabular}{cc}
        \begin{minipage}[t]{0.49\hsize}\centering\includegraphics[width=\linewidth]{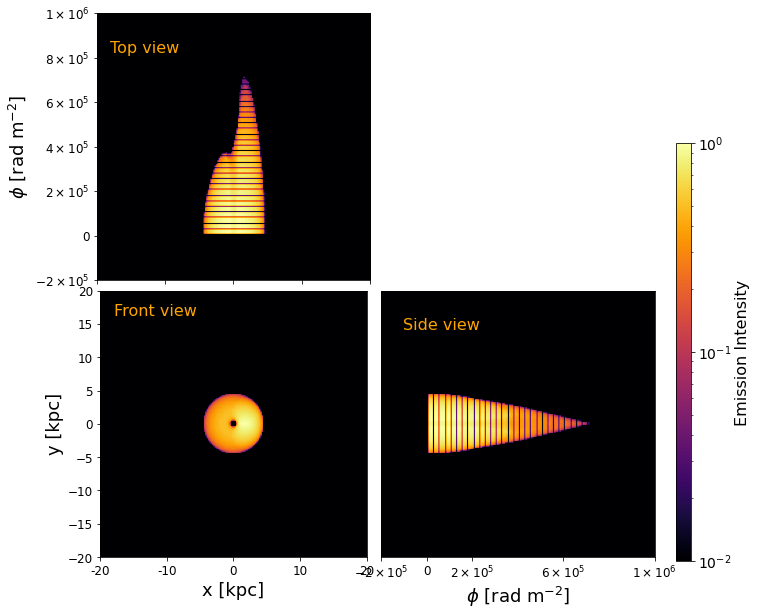}\end{minipage} &
        \begin{minipage}[t]{0.49\hsize}\centering\includegraphics[width=\linewidth]{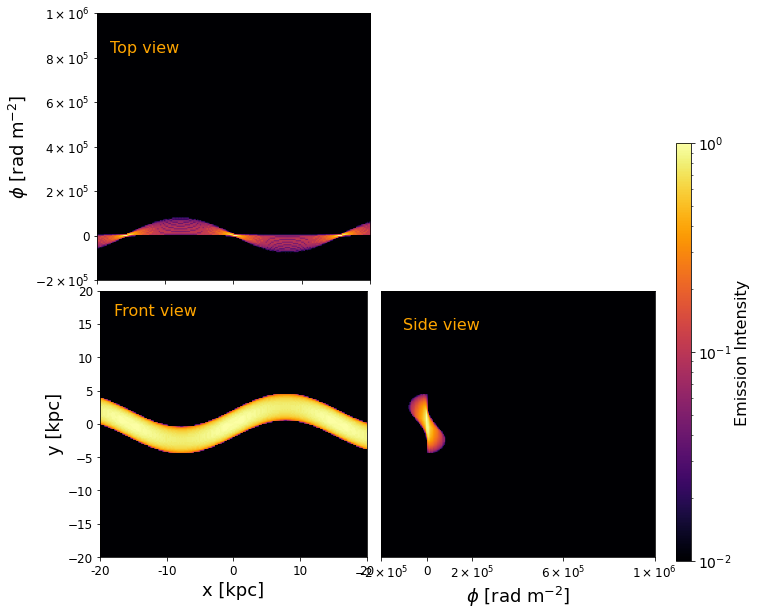}\end{minipage}
    \end{tabular}
    \caption{Three-view visualization of the projected synthetic emission and Faraday structure for the fiducial model, same as Figure~\ref{fig:3views}, but for the two limiting viewing geometries: pole-on $\theta=0$ (left) and plane-of-sky $\theta=\pi/2$ (right).\\
    Alt text: Projected jet emission for pole-on and plane-of-sky viewing geometries.}
    \label{fig:appendix-3views}
\end{figure}

Figure~\ref{fig:appendix-3views} shows the three-view visualization for the two cases. In the pole-on case ($\theta=0$), the projected emission on the sky plane is nearly circular, and the Faraday depth reaches large positive values because the line-of-sight field is dominated by the single-signed axial component $B_{\parallel}$. In the plane-of-sky case ($\theta=\pi/2$), the projected emission traces the full sinusoidal helix, while the Faraday-depth distribution is narrow and symmetric about $\phi=0$: with the jet axis in the plane of the sky the axial field does not contribute to $B_{\parallel,\mathrm{los}}$, so the line-of-sight field is set purely by the azimuthally winding component, which changes sign symmetrically along the jet.

\begin{figure}[tb]
    \centering
    \includegraphics[width=0.33\linewidth]{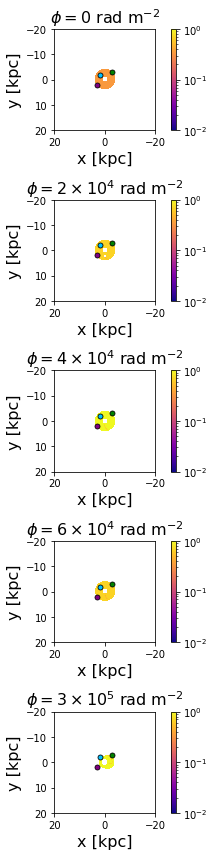}
    \hspace{0.04\linewidth}
    \includegraphics[width=0.33\linewidth]{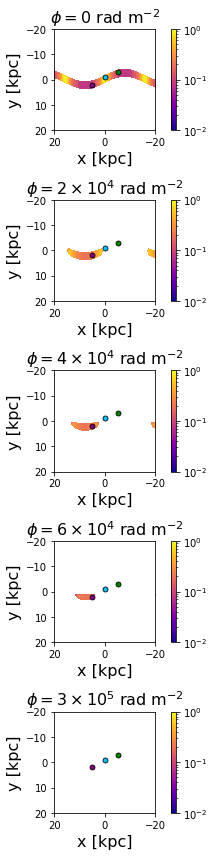}
    \caption{FDF maps at several Faraday-depth slices for the fiducial model, same as Figure~\ref{fig:FDF-abs-image}, but for pole-on $\theta=0$ (left) and plane-of-sky $\theta=\pi/2$ (right).\\
    Alt text: FDF maps for pole-on and plane-of-sky viewing geometries.}
    \label{fig:appendix-fdfmap}
\end{figure}

\begin{figure}[tb]
    \centering
    \begin{tabular}{cc}
        \includegraphics[width=0.46\linewidth]{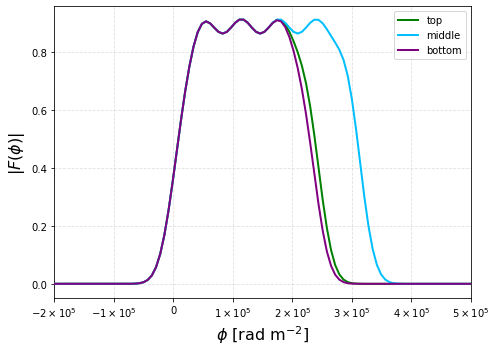} &
        \includegraphics[width=0.46\linewidth]{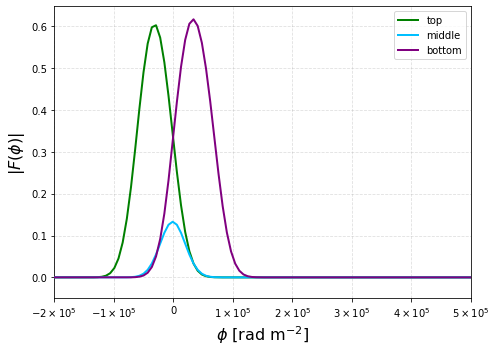}\\
        \includegraphics[width=0.46\linewidth]{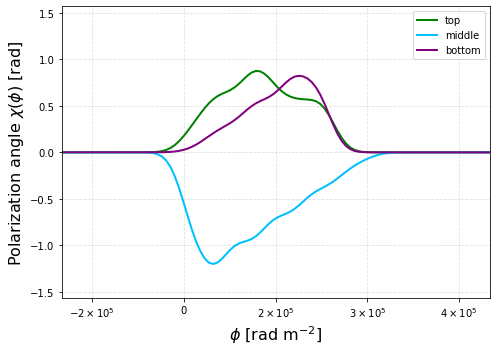} &
        \includegraphics[width=0.46\linewidth]{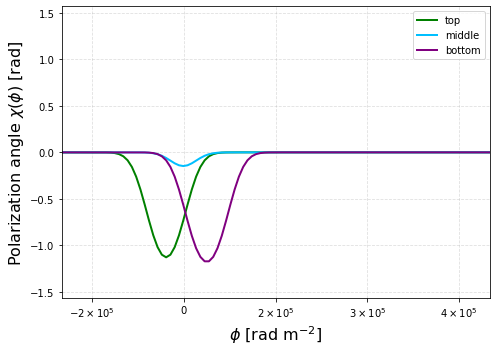}\\
    \end{tabular}
    \caption{Absolute value of the FDF $|F(\phi)|$ (top row) and polarization angle $\chi(\phi)$ (bottom row) along the three representative lines of sight (top, middle, and bottom of the helix), for pole-on $\theta=0$ (left column) and plane-of-sky $\theta=\pi/2$ (right column). For $\theta=\pi/2$ both quantities are symmetric with respect to the sign of $\phi$.\\
    Alt text: FDF amplitude and polarization angle for pole-on and plane-of-sky cases.}
    \label{fig:appendix-fdf}
\end{figure}

Figures~\ref{fig:appendix-fdfmap} and \ref{fig:appendix-fdf} confirm this behavior. For $\theta=\pi/2$, the top and bottom lines of sight show approximately mirrored Faraday-depth distributions, with the top profile extending toward negative $\phi$ and bottom profile toward positive $\phi$, while the middle profile remains close to $\phi=0$, as expected from the symmetry of the plan of sky geometry. It should be noted that, although the Faraday depth distribution is symmetric about $\phi=0$, this symmetry is not apparent in Figure \ref{fig:appendix-fdfmap} because only non-negative Faraday depth slices are shown in the FDF maps. This symmetric limit provides a reference point for interpreting the intermediate inclinations of Section~\ref{sec:orientation}: the residual asymmetry seen there, including the predominantly positive Faraday depths at $\theta=3\pi/8$ (Figure~\ref{fig:3views-inclination}), reflects the finite line-of-sight component of the axial field that persists at all $\theta<\pi/2$.

%\bibliographystyle{abbrvnat}
%\bibliographystyle{}
%\bibliography{references}

\begin{thebibliography}{}

\bibitem[Anderson et al.(2022)]{anderson2022spiderweb}
  Anderson, C. S., Carilli, C. L., Tozzi, P., Miley, G. K., Borgani, S., Clarke, T., Di Mascolo, L., Liu, A., Mroczkowski, T., Pannella, M., et al.\ 2022, ApJ, 937, 45

\bibitem[Baghel et al.(2024)]{baghel2024kpc}
  Baghel, J., Kharb, P., Hovatta, T., Gulati, S., Lindfors, E., \& Silpa, S.\ 2024, MNRAS, 527, 672 

\bibitem[Blandford \& Payne(1982)]{blandford1982hydromagnetic}
  Blandford, R. D., \& Payne, D. G.\ 1982, MNRAS, 199, 883

\bibitem[Blandford \& Znajek(1977)]{blandford1977electromagnetic}
  Blandford, R. D., \& Znajek, R. L.\ 1977, MNRAS, 179, 433

\bibitem[Brentjens \& de Bruyn(2005)]{brentjens2005faraday}
  Brentjens, M. A., \& de Bruyn, A. G.\ 2005, A\&A, 441, 1217

\bibitem[Burn(1966)]{burn1966depolarization}
  Burn, B. J.\ 1966, MNRAS, 133, 67

\bibitem[Carrasco-Gonz\'alez et al.(2010)]{carrasco2010magnetized}
  Carrasco-Gonz\'alez, C., Rodr\'iguez, L. F., Anglada, G., Mart\'i, J., Torrelles, J. M., \& Osorio, M.\ 2010, Sci, 330, 1209

\bibitem[Echibur\'u-Trujillo \& Dexter(2025)]{echiburu2025revealing}
  Echibur\'u-Trujillo, C., \& Dexter, J.\ 2025, ApJ, 985, 260

\bibitem[Gelles et al.(2025)]{gelles2025signatures}
  Gelles, Z., Chael, A., \& Quataert, E.\ 2025, ApJ, 981, 204

\bibitem[Goddi et al.(2025)]{goddi2025first}
  Goddi, C., Carlos, D. F., Crew, G. B., Matthews, L. D., Messias, H., Mus, A., Mart\'i-Vidal, I., Albentosa-Ru\'iz, E., De Laurentis, M., Liuzzo, E., et al.\ 2025, A\&A, 699, A265

\bibitem[Gustafsson et al.(2025)]{gustafsson2025direction}
  Gustafsson, V., Br\"uggen, M., Tasse, C., En\ss lin, T., O'Sullivan, S., \& de Gasperin, F.\ 2025, arXiv:2504.00141

\bibitem[Heald et al.(2020)]{heald2020magnetism}
  Heald, G., Mao, S. A., Vacca, V., Akahori, T., Damas-Segovia, A., Gaensler, B. M., Hoeft, M., et al.\ 2020, Galaxies, 8, 53

\bibitem[Horellou \& Fletcher(2014)]{horellou2014magnetic}
  Horellou, C., \& Fletcher, A.\ 2014, arXiv:1401.4152

\bibitem[Jerrim et al.(2024)]{jerrim2024faraday}
  Jerrim, L. A., Shabala, S. S., Yates-Jones, P. M., Krause, M. G. H., Turner, R. J., Anderson, C. S., Stewart, G. S. C., Power, C., \& Rodman, P. E.\ 2024, MNRAS, 531, 2532

\bibitem[Jerrim et al.(2025)]{jerrim2025braise}
  Jerrim, L., Shabala, S., Yates-Jones, P., Krause, M., Turner, R., Stewart, G., \& Power, C.\ 2025, PASA, 42, e136

\bibitem[Kim et al.(2024)]{kim2024magnetic}
  Kim, D. E., Di Gesu, L., Liodakis, I., Marscher, A. P., Jorstad, S. G., Middei, R., Marshall, H. L., Pacciani, L., Agudo, I., Tavecchio, F., et al.\ 2024, A\&A, 681, A12

\bibitem[Kramer et al.(2025)]{kramer2025probing}
  Kramer, J. A., M\"uller, H., R\"oder, J., \& Ros, E.\ 2025, A\&A, 697, A66

\bibitem[Livingston et al.(2025)]{livingston2025helical}
  Livingston, J. D., Nikonov, AS., Dzib, S. A., Debbrecht, L. C., Kovalev, Y. Y., Lisakov, M. M., MacDonald, N. R., Paraschos, G. F., R\"oder, J., \& Wielgus, M.\ 2025, A\&A, 695, A260

\bibitem[Meenakshi et al.(2024)]{meenakshi2024comparative}
  Meenakshi, M., Mukherjee, D., Bodo, G., Rossi, P., \& Harrison, C. M.\ 2024, MNRAS, 533, 2213

\bibitem[Orienti et al.(2024)]{orienti2024high}
  Orienti, M., Siemiginowska, A., D’Ammando, F., \& Migliori, G.\ 2024, A\&A, 687, A287

\bibitem[Park et al.(2026)]{park2026helical}
  Park, J., Takahashi, K., Toma, K., Hada, K., Nakamura, M., Pu, H.-Y., Asada, K., Ho, P. T. P., Kino, M., Kawashima, T., et al.\ 2026, ApJ, 996, L22

\bibitem[Pasetto et al.(2021)]{pasetto2021reading}
  Pasetto, A., Carrasco-Gonz\'alez, C., Gomez, J. L., Mart\'i, J.-M., Perucho, M., O’Sullivan, S. P., Anderson, C., D\'iaz-Gonz\'alez, D. J., Fuentes, A., \& Wardle, J.\ 2021, ApJ, 923, L5

\bibitem[Peng et al.(2024)]{peng2024faraday}
  Peng, S., Lu, R.-S., Goddi, C., Krichbaum, T. P., Li, Z., Liu, R.-Y., Kim, J.-Y., Nakamura, M., Yuan, F., Chen, L., et al.\ 2024, ApJ, 975, 103

\bibitem[Pushkarev et al.(2023)]{pushkarev2023mojave}
  Pushkarev, A. B., Aller, H. D., Aller, M. F., Homan, D. C., Kovalev, Y. Y., Lister, M. L., Pashchenko, IN., Savolainen, T., \& Zobnina, D. I.\ 2023, MNRAS, 520, 6053

\bibitem[Rodr\'iguez-Kamenetzky et al.(2025)]{rodriguez2025helical}
  Rodr\'iguez-Kamenetzky, A., Pasetto, A., Carrasco-Gonz\'alez, C., Rodr\'iguez, L. F., G\'omez, J. L., Anglada, G., Torrelles, J. M., Gomes, N. R. C., Vig, S., \& Mart\'i, J.\ 2025, ApJ, 978, L31

\bibitem[Rudnick et al.(2024)]{rudnick2024pseudo}
  Rudnick, L., Anderson, C., Cotton, W. D., Pasetto, A., Alexander, E. L., \& Tahani, M.\ 2024, MNRAS, 535, 2115

\bibitem[Sakemi et al.(2025)]{sakemi2025three}
  Sakemi, H., Chibueze, J. O., Cotton, W. D., Parekh, V., Ohmura, T., Machida, M., Igarashi, T., Akahori, T., Akamatsu, H., Nakanishi, H., et al.\ 2025, ApJ, 992, 14

\bibitem[Takahashi(2023)]{2023PASJ...75S..50T}
  Takahashi, K.\ 2023, PASJ, 75, S50

\bibitem[Toscano et al.(2025)]{toscano2025helical}
  Toscano, T., Molina, S. N., G\'omez, J. L., Zeng, A.-L., Dahale, R., Cho, I., Moriyama, K., Wielgus, M., Fuentes, A., Foschi, M., et al.\ 2025, A\&A, 698, A210

\bibitem[Tsunetoe et al.(2025)]{tsunetoe2025polarization}
  Tsunetoe, Y., Narayan, R., \& Ricarte, A.\ 2025, American Astronomical Society Meeting Abstracts, 245, 256--03

\bibitem[Yang \& Jiang(2025)]{yang2025spectrum}
  Yang, S., \& Jiang, Y.-G.\ 2025, PASP, 137, 104101

\end{thebibliography}

\end{document}